\documentclass[a4paper,fleqn]{cas-dc}

\usepackage[numbers]{natbib}
\usepackage{hyperref}
\usepackage{xurl}
\usepackage{caption}

\def\tsc#1{\csdef{#1}{\textsc{\lowercase{#1}}\xspace}}
\tsc{WGM}
\tsc{QE}

\begin{document}
\let\WriteBookmarks\relax
\def\floatpagepagefraction{1}
\def\textpagefraction{.001}

\shorttitle{RansomTrack for Ransomware Detection}    

\shortauthors{Caliskan B. et al.}  

\title [mode = title]{RansomTrack: A Hybrid Behavioral Analysis Framework for Ransomware Detection}  

\tnotemark[1,2] 

\tnotetext[1]{This study was supported in part by the Scientific Research Projects Department of Istanbul Technical University (Project ID Number: 45437 and Project Code: MAB-2024-45437) and EUREKA cluster ITEA project VESTA  which is also supported by TUBITAK (The Scientific and Technological Research Council of Türkiye).} 
\tnotetext[2]{All of the codes developed for this project and the collected RansomTrack dataset are published in our Github Repository. \url{https://github.com/bcandroid/A-Hybrid-Behavioral-Analysis-Dataset-for-Ransomware-and-Benignware-Detection}}

%

\author[1,3]{BUSRA CALISKAN}[
                        orcid=0009-0006-2595-4873]
\cormark[1]
\ead{caliskanb21@itu.edu.tr}

\affiliation[1]{organization={Department of Computer Engineering, Istanbul Technical University},
                city={Istanbul},
                country={Türkiye}}

\author[2]{IBRAHIM GULATAS}[
                        orcid=0000-0002-0804-3588]
\cormark[2]
\ead{igulatas@hotmail.com}
\affiliation[2]{organization={Department of Research and Development, Turkish Naval Forces},
                city={Istanbul},              
                country={Türkiye}}

\author[3]{H. HAKAN KILINC}[
                        orcid=0000-0001-6610-6324]
\ead{hakan.kilinc@orioninc.com}

\affiliation[3]{organization={Department of Research and Development, Orion Innovation},
                city={Istanbul},
               country={Türkiye}}

\author[1]{A. HALIM ZAIM}[
                        orcid=0000-0002-0233-064X]

\ead{azaim@itu.edu.tr}

\cortext[cor1]{Principal Corresponding author}
\cortext[cor2]{Corresponding author}

\begin{abstract}
Ransomware poses a serious and fast-acting threat to critical systems, often encrypting files within seconds of execution. Research indicates that ransomware is the most reported cybercrime in terms of financial damage, highlighting the urgent need for early-stage detection before encryption is complete. In this paper, we present RansomTrack, a hybrid behavioral analysis framework to eliminate the limitations of using static and dynamic detection methods separately. Static features are extracted using the Radare2 sandbox, while dynamic behaviors such as memory protection changes, mutex creation, registry access and network activity are obtained using the Frida toolkit. Our dataset of 165 different ransomware and benign software families is publicly released, offering the highest family-to-sample ratio known in the literature. Experimental evaluation using machine learning models shows that ensemble classifiers such as XGBoost and Soft Voting achieve up to 96\% accuracy and a ROC-AUC score of 0.99. Each sample analyzed in 9.1 seconds includes modular behavioral logging, runtime instrumentation, and SHAP-based interpretability to highlight the most influential features. Additionally, RansomTrack framework is able to detect ransomware under 9.2 seconds. Overall, RansomTrack offers a scalable, low-latency, and explainable solution for real-time ransomware detection.
\end{abstract}


\begin{highlights}
\item A hybrid ransomware detection framework called RansomTrack using static and dynamic behavioral features.
\item RansomTrack enables low latency ransomware detection below 9.2 seconds by using Frida and Radare2.
\item RansomTrack offers SHAP-based interpretability to highlight indicators of critical behavior.
\item The public dataset includes 165 ransomware families and 2410 samples.
\end{highlights}


\begin{keywords}
 Ransomware Detection\sep Static Analysis\sep Dynamic Analysis\sep Behavior Analysis\sep Hybrid Model\sep Pearson Correlation\sep
 
\end{keywords}

\maketitle

\section{Introduction}\label{sec:introduction}

Ransomware has emerged as one of the most critical cybersecurity threats due to its ability to rapidly encrypt data, disrupt essential services, and demand payment in exchange for decryption. Since the appearance of the first known ransomware, AIDS Trojan, in 1989, the frequency and impact of ransomware attacks have escalated significantly, especially after 2020. Recent high-profile ransomware incidents have demonstrated the disruptive and costly nature of such attacks across various critical sectors. For instance, the 2021 Colonial Pipeline attack caused a six-day disruption in fuel delivery across the U.S. East Coast~\cite{bellamkonda2021colonial}, while the Kaseya supply-chain attack affected approximately 1,500 organizations and involved a 70 million dollar ransom demand—the highest publicly disclosed to date~\cite{cisa2021kaseya}. In the healthcare sector, Ireland’s Health Service Executive canceled over 200,000 medical appointments and incurred more than 100 million euros in damages due to a Conti infection~\cite{oconnor2021hse}. In 2022, Costa Rica declared a national emergency after 27 government agencies were disabled by ransomware, prompting 25 million dollars in aid from the United States~\cite{gonzalez2022costarica}.

According to the FBI’s Internet Crime Complaint Center (IC3) 2024 report, there were 3,156 ransomware complaints in that year—a 9\% increase over 2023—with total reported losses reaching 12.47 million dollars~\cite{ic3ransomware2024}.  These developments emphasize the urgency of designing low-latency, resilient detection mechanisms against increasingly evasive ransomware.

Another notable aspect is that the Windows operating systems are prime targets for ransomware campaigns due to their widespread use, rich API surface, and architectural complexity. Threat intelligence reports indicate that more than 90\% of ransomware incidents occur on Windows platforms~\cite{statista2022osshare,halcyon2023execs}.

Despite significant advancements in analysis methods, ransomware continues to challenge both static and dynamic detection techniques.

\begin{itemize}
\item \textbf{Static analysis is hindered by obfuscation, packing, and control flow manipulation} In static analysis, where binaries are inspected without execution, key limitations arise due to common techniques such as obfuscation, packing, and control-flow manipulation. Many ransomware families generate large and irregular basic blocks that impair disassembly and control-flow reconstruction \cite{kang2023survey}. Obfuscation techniques such as control-flow flattening, opaque predicates, junk code insertion, and instruction substitution further degrade the accuracy of static analysis \cite{okhravi2020software}. Moreover, ransomware is frequently packed or encrypted using tools like UPX or Themida, concealing the true payload until runtime, thereby bypassing signature-based detection unless unpacking is explicitly performed \cite{bilge2019dissecting}.
\item \textbf{Dynamic analysis can be evaded through sandbox-aware behaviors, delayed execution, or interaction-triggered payloads}. Dynamic analysis, though capable of revealing runtime behavior, is equally susceptible to evasion strategies. Environment-aware ransomware variants routinely detect virtualized or sandboxed environments by probing for known registry keys, drivers, and timing discrepancies \cite{olaimat2021taxonomy}. Furthermore, delayed execution techniques, including long sleep calls or busy loops, aim to exceed the analysis time window of automated sandboxes \cite{jiang2021exploring}. Some samples rely on user interaction triggers such as mouse movement or keyboard input, remaining dormant until such events are observed, effectively bypassing fully automated dynamic analysis systems \cite{mathew2020delayed}. 
\end{itemize}  

\subsection{Motivation} 

These evasive strategies render many single-modality solutions ineffective. Ransomware actively employs anti-analysis mechanisms to evade detection in both static and dynamic techniques.  These limitations emphasize the clear need for hybrid, explainable, and real-time detection systems

In this work, we present RansomTrack, a hybrid ransomware detection framework and dataset. The proposed framework uses the Radare2~\cite{radare22020} framework for reverse engineering and analyzing binaries to extract static features and Frida~\cite{frida2022}, a lightweight runtime instrumentation toolkit, to extract dynamic features. This combination allows for robust and repeatable profiling across various ransomware families.

\textbf{The key contributions of this study include:}
\begin{itemize}
    \item A publicly available hybrid ransomware dataset combining opcode-level static and dynamic features, including API calls and page memory protection flags, of 165 ransomware families.
    \item A low-latency detection pipeline (9.1 sec/sample) with real-time behavioral tracing.
    \item Modular instrumentation using Radare2 and Frida for reproducibility and resilience.
    \item SHAP-based interpretability to highlight the most impactful behavioral features.
\end{itemize}

The remainder of this paper is organized as follows. Section~\ref{sec:related_work} reviews related literature, focusing on existing analysis frameworks, tools, and datasets. Section~\ref{sec:framework} presents the proposed hybrid feature extraction pipeline and dataset construction methodology. Section~\ref{sec:results} details the experimental setup, model configurations, and performance evaluation. Section~\ref{sec:discussion} interprets the findings and compares classifier behavior. Finally, Section~\ref{sec:conclusion} concludes the paper and discusses future directions.

\section{Related Work}
\label{sec:related_work}

Ransomware detection has received considerable attention in recent years due to the increasing complexity and frequency of attacks. To address this challenge, researchers have proposed a range of analysis techniques that leverage static, dynamic and hybrid analysis paradigms, each offering complementary insights into executable behavior. Static analysis focuses on structural features such as sequences of opcode, control flow graphs and API imports without executing the binary. In contrast, dynamic analysis captures real-time runtime interactions, including API calls, memory protection changes and registry manipulations. To overcome the individual limitations of these methods, such as obfuscation resistance in static analysis or evasion tactics in dynamic environments, researchers have increasingly focused on hybrid frameworks. Hybrid frameworks have shown promising results in improving detection accuracy and resilience to evasion. In this section, prominent tools and datasets supporting these methodologies are reviewed, with an emphasis on static feature extraction (via Radare2), dynamic behavior analysis (via Frida), and hybrid frameworks and dataset building efforts in recent literature.

\subsection{Static Analysis with Radare2}

Among existing static analysis tools, Radare2 stands out due to its scriptability, performance, and open source accessibility. Unlike commercial tools such as IDA Pro~\cite{idapro}, Radare2 supports large-scale batch processing and seamless automation through its Python interface r2pipe.

Previous work has successfully used Radare2 for opcode extraction and control flow graph generation ~\cite{mester, koppanati, gulatas}. For example, \cite{koppanati} used Radare2 to extract blocks of opcode surrounding API calls and transform them into opcode “sentences” for BERT-based classification. Their approach achieved 99.93\% accuracy without false positives, outperforming traditional n-gram and full binary models. Similarly, \cite{gulatas} applied Radare2 to MIPS-based IoT malware to classify opcodes into functional categories (e.g., memory, branching, logic) and achieved high accuracy with reduced feature dimensionality while preserving behavioral significance. The resulting features provided high accuracy classification (up to 99\%) and reduced detection latency by a factor of 7.2$\times$, highlighting the utility of Radare2 in resource-constrained scenarios.

In the field of ransomware detection, opcode-level patterns offer important static signals such as encryption routines, control flow obfuscation, or anti-debugging techniques. The proposed framework extracts structured opcode sequences from 32-bit Windows PE files using the Radare2 tool. This method provides a feature representation that is both lightweight and informative, resistant to common evasion tactics such as API import obfuscation or PE header tampering.

\subsection{Dynamic Analysis with Frida}

For observing malware behavior at runtime and gaining insight into system interactions that cannot be statically inferred, the Frida toolkit stands out. Frida is a lightweight and codeable instrumentation toolkit that has been widely adopted in dynamic malware analysis due to its low overhead, cross-platform support, and real-time hook capabilities. Unlike full sandbox solutions, Frida operates at the API level, allowing precise monitoring of function calls, arguments, return values, and memory events.

Recent research has demonstrated Frida's effectiveness in catching evasive behavior.  ~\cite{soriano2023} investigated how advanced malware can evade Frida by detecting injected memory artifacts or restoring original function initializations, highlighting both Frida's power and limitations in stealth analysis.

~\cite{zhou2024} used Frida in their APInspector framework to monitor Windows API calls at runtime. They defined \texttt{onEnter} and \texttt{onLeave} callbacks to capture instrumentation scripts, input parameters (i.e., file paths, registry keys, IPs), and return values. These features formed the input for a hybrid model combining a Hierarchical Attention Network (HAN) for API sequences and an MLP for parameter values, achieving 97.5\% accuracy.

Drabent et al.\cite{drabent2024} implemented Frida on Android ransomware to reveal obfuscated C\&C traffic and injected payload logic. In another Android malware analysis, Bae et al.~\cite{bae2022} applied Frida as part of the AMS sandbox system to dynamically manipulate application behavior and avoid emulation-based detection. Frida is used to modify runtime values such as IMEI and IMSI to emulate real devices.

In our framework, Frida is used to monitor dynamic behavioral indicators related to ransomware execution, including memory protection changes, mutex creation, registry changes, file operations, and network activity. By injecting custom JavaScript hooks into 32-bit Windows PE processes, our solution captures detailed behavioral logs in real-time while maintaining low execution latency. This enables early-stage ransomware detection before encryption routines are fully executed.

\subsection{Framework and Dataset Landscape}

\begin{table*}[H]
\caption{Summary of the Ransomware and Benignware Datasets}
\centering
\renewcommand{\arraystretch}{1.15}
\begin{scriptsize}
\begin{tabular}{|>{\centering\arraybackslash}p{1.5cm}|>{\centering\arraybackslash}p{1.8cm}|>{\centering\arraybackslash}p{1.9cm}|>{\centering\arraybackslash}p{1.7cm}|>{\centering\arraybackslash}p{1.9cm}|>{\centering\arraybackslash}p{2.1cm}|>{\centering\arraybackslash}p{1.6cm}|>{\centering\arraybackslash}p{2cm}|}
\hline
\textbf{Aspect} & \textbf{RansomTrack (2025)} & \textbf{RansomFormer (2024) ~\cite{alzahrani2025ransomformer}} & \textbf{RanSMAP (2024) ~\cite{Hirano}} & \textbf{Ransomware / Benignware System Calls (2023) ~\cite{dib}} & \textbf{Dynamic Feature Dataset (2022) ~\cite{Herrera}} & \textbf{RanSAP (2021) ~\cite{hirano2022ransap}} & \textbf{CICAndMal2017 (2017) ~\cite{lashkari}} \\
\hline
\textbf{Feature Types} & Hybrid (Static + Dynamic) & Hybrid (Static + Dynamic) & Dynamic only: Memory and Storage & Dynamic only: System calls & Dynamic only: Behavioral features & Dynamic only: Storage I/O patterns & Hybrid (APK + Network behavior) \\
\hline
\textbf{Static Features} & Opcodes & Raw PE byte sequences, Imported API functions & \textit{Not included} & \textit{Not included} & \textit{Not included} & \textit{Not included} & APKs provided (permissions, manifest analyzable) \\
\hline
\textbf{Static Feature Tools} & Radare2 & \texttt{pefile} Python module & \textit{Not applicable} & \textit{Not applicable} & \textit{Not applicable} & \textit{Not applicable} & ApkTool, VirusTotal (implied) \\
\hline
\textbf{Dynamic Features} & API Calls and Page Protection Flags  & API call sequences & Memory access patterns, Storage sector ops & System call traces  & File ops, API usage, Registry, Services & Disk sector traces, Time-series I/O & Network traffic, Flow stats, Battery/API logs \\
\hline
\textbf{Dynamic Collection Tools} & Frida & Cuckoo Sandbox 3 & BitVisor Hypervisor, UDP logging & API Monitor (manual execution) & Cuckoo (JSON), Feature pipeline & Custom hypervisor tracer, Disk I/O monitor & Tcpdump, CICFlowMeter, Appmon, ADB tools \\
\hline
\textbf{Operating System and File Types} & Win 10 Pro 32-Bit, .exe files & Win 7/8/10/11,.exe files & Windows 10,.exe files & Windows 10 32-bit, .exe files & Win XP/7/10, .exe files & Windows 10, .exe files & Android 5.x--6.x, APKs \\
\hline
\textbf{Execution Environment} & Oracle VirtualBox & Cuckoo-based sandbox VMs & Bare-metal w/ BitVisor & Manual VM exec w/ INetSim & Cuckoo sandbox (5 platforms) & Virtualized w/ hypervisor monitor & Real Android phones, WiFi hotspot \\
\hline
\textbf{Sample Sources} & Ransomware: MalwareBazaar~\cite{malwarebazaar},~\cite{moreira2023ransomware}; Benignware: ~\cite{moreira2023ransomware}, ~\cite{dikedataset}, ~\cite{benignnet} & VirusTotal, MalwareBazaar, VirusShare & ANY.RUN, VirusTotal, real apps & AnyRun, VirusShare, HybridAnalysis, MajorGeeks & VirusTotal, HybridAnalysis, TheZoo & Public ransomware, browsers & CIC, APKMirror, VirusTotal \\
\hline
\textbf{Sample Counts} & 1205 ransomware and benignware & 2,000 ransomware and benignware & 6 ransomware, 6 benign, 7 variants & 270 ransomware, 270 benign & 1,000 ransomware, 1,000 benign & 7 ransomware, 5 benign, 21 variants & 10,854 total APKs, 429 malware, 5,065 benign executed \\
\hline
\textbf{Ransomware Families} & 165 families & 161 families & 7 families & 12 families (e.g., Petya, Xorist) & 20 families (encryptors/lockers) & 7 families (e.g., Locky, CryptoLocker) & 10 families (e.g., Koler, Svpeng, Charger) \\
\hline
\textbf{ML/DL Model Used} & Soft voting, RF, DT, KNN, MLP, LR & Dual-stream Transformer & 1D CNN + LSTM & RF, MLP, SVM, LightGBM, XGBoost & RF, GBT, Naive Bayes, NN & Logistic Regression, SVM, LightGBM & RF, KNN, J48 Decision Tree \\
\hline
\textbf{Performance Metrics} & Acc: 0.96, Rec: 0.96, Prec: 0.97 & Acc: 99.50\%, Prec: 99.67\%, Rec: 99.33\% & F1-score $\approx$ 94.3\% & Acc: 99.81\%, Prec: 99.64\%, Rec: 100\%, F1: 99.82\% & Acc $>$ 0.99, F1 $\approx$ 0.99 (10-fold CV) & F1 $\approx$ 92.0\%, Prec 91.7\% & Acc: 85--88\%, Category F1: $\sim$50\%, Family F1: $\sim$27\% \\
\hline
\end{tabular}
\label{tab:ransom_datasets_comparison}
\end{scriptsize}
\end{table*}

In this section, we will review the prominent works in ransomware detection and especially the datasets they use and present a comparison with our work. The development of robust ransomware detection systems is highly dependent on the availability of high-quality datasets that capture a variety of malware behaviors. The existing public datasets differ in terms of feature types (static, dynamic or hybrid), data collection strategies and ransomware family coverage.

One prominent hybrid framework is the \textit{RansomFormer} framework ~\cite{alzahrani2025ransomformer}, which integrates raw static byte streams and API import tables with dynamic API call sequences extracted from ransomware and benignware PE files using the Cuckoo Sandbox. Their dual-stream Transformer-based architecture achieves high classification performance—99.25\% with static features alone and 99.50\% when hybrid inputs are fused via a cross-attention mechanism. The dataset used to test the framework covers 161 ransomware families and is widely used in multimodal malware detection studies. However, several limitations exist: the raw binaries are not shared, static features are susceptible to packing and obfuscation, dynamic traces can be evaded by sandbox-aware or time-delayed malware, the hybrid subset is relatively small due to execution constraints, and the dataset lacks evaluation against adversarial threats such as API padding or poisoning attacks.

Another noteworthy dataset is RanSMAP~\cite{Hirano}, which is designed to capture low-level behavioral signals by tracing both memory and storage access patterns at the hypervisor level. Unlike traditional OS-level datasets that rely on API calls or system logs, RanSMAP leverages BitVisor to record fine-grained memory page events and disk sector I/O, thereby enabling robust ransomware detection even under API obfuscation or sandbox-evasion techniques. The dataset includes over 11,000 trace files derived from multiple hardware configurations and execution contexts, covering six ransomware families and several benign utilities. Its key contribution lies in bypassing semantic obfuscation by operating below the OS layer, while offering a reproducible benchmark for deep learning-based analysis. However, its use of low-level features introduces a semantic gap, lacks process-level context, and imposes deployment challenges due to the need for bare-metal hypervisor instrumentation.

Dib et al.~\cite{dib} created a dataset from scratch and built machine learning models using eight different algorithms for ransomware/benignware system calls. The dataset contains dynamic behavioral traces collected from 270 ransomware and 270 benign samples executed in a controlled Windows 10 environment. The dataset covers 12 ransomware families, including CryptoLocker, Petya, and TeslaCrypt. API/system calls were monitored via API Monitor during manual execution, with INetSim used to simulate internet activity.To overcome the bias of conventional feature selection pipelines that often favor benignware activity, the authors propose a hybrid strategy combining Spearman and Kendall correlation filtering, Permutation Feature Importance (PFI), and a TF-IDF-inspired method called FICFR (Feature Importance based on Call Frequency and References). This approach effectively highlights ransomware-specific system calls.The best-performing model, a Multi-Layer Perceptron (MLP), achieved high scores using only 16 selected features—demonstrating high performance with low-dimensional input.Limitations include challenges in capturing live ransomware behaviors due to inactive C2 servers and the computational overhead of correlation-based feature evaluation on large dimensions.

In the study by Herrera-Silva and Hernández-Álvarez ~\cite{Herrera}, a dataset was produced and analyzed through 4 different models using this dataset. This dataset called the Dynamic Feature Dataset comprises 2,000 instances extracted from Cuckoo sandbox reports, executed across five Windows versions. From over 326 raw behavioral features, the authors selected the top 50 most relevant indicators (e.g., encryption behavior, API misuse, service creation) using correlation analysis. The dataset supports reproducibility and variant-resilient ML detection. However, it excludes static features, lacks raw execution logs, and requires manual JSON-to-feature conversion, which limits scalability. Despite these constraints, it contributes a practical benchmark for behavior-based ransomware detection and early-stage ML pipeline evaluation.

Another dataset that Hirano et al. presented 3 years before the RanSMAP dataset~\cite{Hirano} is the RanSAP dataset~\cite{hirano2022ransap}. It provides a novel dynamic analysis dataset that captures low-level storage I/O patterns using a thin hypervisor. By tracing disk sector reads/writes, it enables early ransomware detection (within 30 seconds) through entropy and access frequency patterns. RanSAP is the first open dataset to monitor disk activity at the hypervisor level, offering resilience against API-level evasion. However, it lacks static features and semantic OS-level context, and its detection accuracy decreases under full-disk encryption. The dataset remains limited in size and requires manual preprocessing of CSV traces into feature vectors.

Finally, CICAndMal2017 dataset, proposed by Lashkari et al.~\cite{lashkari}, introduces a real-device-based Android malware benchmark that emphasizes capturing realistic runtime and network behaviors. It includes both benign and malicious Android apps, executed under controlled user-interaction scenarios on real smartphones, with network traffic recorded and labeled across five malware categories. Over 80 network flow features were extracted using CICFlowMeter, enabling robust malware binary classification through machine learning models such as Random Forest, KNN, and Decision Tree. The dataset achieves high accuracy for binary detection.However, it exhibits limitations in fine-grained classification: family-level and category-level detection yielded lower performance due to insufficient per-family samples and overlapping traffic patterns. Additionally, the dataset does not provide pre-extracted static features, and malware labeling relies on the majority Anti-Viruses (AV) vendor consensus, which can introduce inconsistency.

As shown in Table~\ref{tab:ransom_datasets_comparison}, the dataset of our proposed framework, RansomTrack, provides a large-scale, hybrid collection of 2,410 labeled Windows PE files across 165 ransomware families and their benign correspondences. Static features are extracted as opcode arrays using Radare2, while dynamic features are obtained from Frida-based instrumentation to capture fine-grained behavioral traces. Compared to other studies, RansomTrack's differences are as follows.

\begin{itemize}
    \item It has the highest known family-to-sample ratio to maximize behavioral diversity.
    \item It is a modular feature extraction framework with producible tools.
    \item It uses SHAP-based interpretability to increase transparency in classification.
    \item It has real-time applicability with low per-sample latency (9.1 sec).
\end{itemize}

To develop effective detection strategies against modern ransomware, it is essential to understand both the static structure and dynamic behavior of malicious executable files.

\subsection{Tool-Based Ransomware Analysis}

Effective ransomware analysis often requires the orchestration of multiple tools across static, dynamic, network traffic monitoring and memory forensics. These tools not only support feature extraction, but also facilitate reproducibility, automation, and resilience to evasion techniques. Table~\ref{tab:tools} summarizes the tools commonly used in ransomware investigations, categorized according to their analytical role.

\begin{table}[htp]
\centering
\begin{scriptsize}
\caption{Categorization of Tools Used in Ransomware Analysis}
\label{tab:tools}
\begin{tabular}{|>{\centering\arraybackslash}p{3.5cm}|>{\centering\arraybackslash}p{1.1cm}|>{\centering\arraybackslash}p{2.4cm}|}
\hline
\textbf{Tool} & \textbf{Category} & \textbf{Function} \\ \hline
Radare2~\cite{radare22020}, Ghidra~\cite{ghidra2019}, Capstone~\cite{nguyen2014capstone}, PEStudio~\cite{pestudio2023}, IDA Pro ~\cite{idapro} & Static Analysis & Disassembly, opcode extraction \\ \hline
Frida~\cite{frida2022}, ProcMon~\cite{procmon2020}, RegShot~\cite{regshot2019}, ProcDot~\cite{procdot2021} & Dynamic Analysis & API monitoring, registry tracing \\ \hline
Cuckoo Sandbox~\cite{cuckoo2016}, CAPEv2~\cite{cape2020}, Volatility~\cite{volatility2014}, Noriben~\cite{noriben2020}, DumpIt~\cite{comae2020}   & Memory/ Sandbox & Process analysis, memory forensics \\ \hline
YARA~\cite{yara2013}, Wireshark~\cite{wireshark2023}, INetSim~\cite{inetsim2019} & Support Tools & Rule-based detection, traffic emulation \\ \hline
\end{tabular}
\end{scriptsize}
\end{table}

Static analysis tools such as Radare2~\cite{radare22020}, Capstone~\cite{nguyen2014capstone} and Ghidra~\cite{ghidra2019} enable disassembly, control flow graphing and opcode extraction from Windows PE binaries. These tools are essential for revealing structural patterns such as encryption loops or obfuscation structures and for performing PE anomaly detection, debugging and reverse engineering.

Dynamic analysis tools such as Frida~\cite{frida2022}, ProcMon~\cite{procmon2020} and RegShot~\cite{regshot2019} provide runtime visibility into API calls, registry access, mutual exclusion generation, behavior correlation and network communications.

Frameworks such as Cuckoo Sandbox~\cite{cuckoo2016}, CAPEv2, Volatility~\cite{volatility2014} and Noriben~\cite{noriben2020} for memory and sandbox-based analysis, especially for packaged or memory-resident ransomware variants, offer deeper insight into transaction behavior, memory analysis, genetic malware attribution and forensic artifacts.

Finally, support tools such as YARA~\cite{yara2013}, INetSim~\cite{inetsim2019} and Wireshark~\cite{wireshark2023} enable rule-based detection, simulated fake network services and traffic inspection and monitoring, facilitating the study of command and control patterns and anti-analysis strategies.

These tools can be combined to form hybrid analysis pipelines capable of capturing both structural and behavioral indicators of ransomware activity. In our work, Radare2 and Frida are integrated tightly to provide opcode-level disassembly and runtime instrumentation respectively, which forms the core of the RansomTrack feature extraction system.

\section{Proposed Framework}\label{sec:methodology}

To overcome the limitations of single-modality detection systems, we propose a hybrid feature extraction pipeline that combines both static and dynamic behavior indicators. Our framework leverages the strengths of Radare2 for structural opcode-level sharding and Frida for real-time API and Page Memory Protection Flag monitoring during binary execution. This dual-modal approach enables a more comprehensive representation of ransomware behavior while maintaining low computational overhead.

\begin{figure*}[htbp]
  \centering
  \includegraphics[width=0.9\textwidth,keepaspectratio]{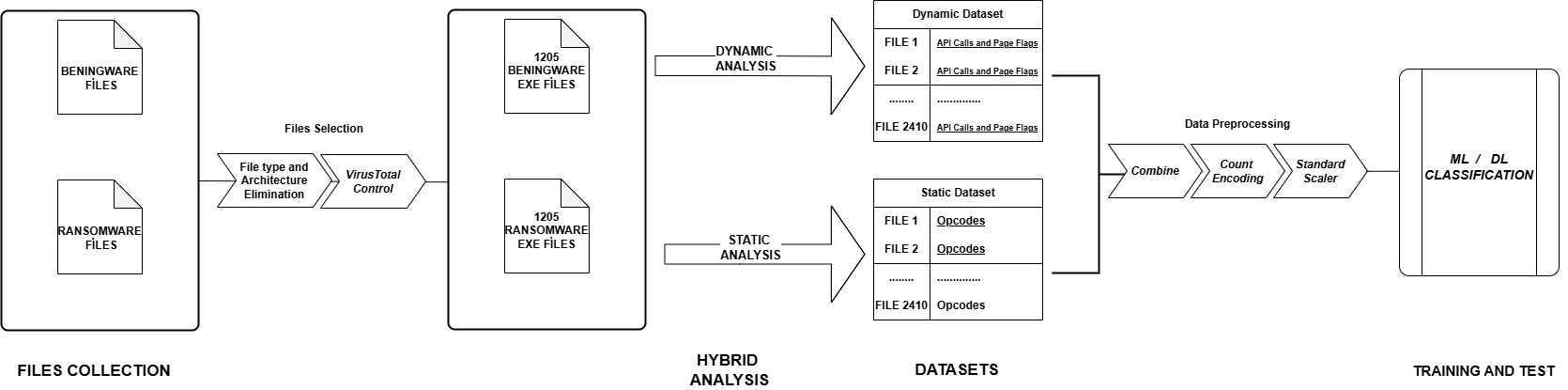}
  \caption{End-to-End Ransomware Detection Pipeline Using Static and Dynamic Features}
  \label{fig:workflow}
\end{figure*}

The proposed RansomTrack framework is developed in four main phases. First of all, ransomware and benignware samples are collected in order to create the dataset. Then, static malware analysis techniques are applied for feature extraction. Following this, dynamic analysis techniques are applied to the same collection in order to generate a hybrid dataset. Finally, various data preprocessing techniques and classification algorithms are applied to the dataset to train the optimum ML model. The general overview of
The RansomTrack framework is shown in Figure ~\ref{fig:workflow} which illustrates the overall pipeline including sample preprocessing, instrumentation, feature extraction, and classifier input generation. 

\subsection{Dataset Collection}

Developing an effective ransomware detection system critically depends on a comprehensive understanding of ransomware behavior. However, publicly available ransomware datasets in the literature remain limited in both quantity and quality. Many existing datasets are outdated and primarily focus on specific ransomware families, often lacking diversity in attack techniques and system environments. More critically, most of these datasets do not include benign software samples, which are essential for training reliable binary classification models. Additionally, collecting representative benignware is a non-trivial task, as it requires careful curation of legitimate software across various system configurations and environments. The absence of standardized benchmarks across studies also complicates fair comparison of detection approaches developed on the same dataset.

The "Portable Executable (PE)" file format is the standard binary format used by the Windows operating system for executables, dynamic link libraries (DLLs), and other system-level components. It encapsulates the structural metadata required by the OS loader to map the binary into memory and initiate execution~\cite{microsoft_pe, wikipedia_pe}. In this study, we specifically focus on 32-bit Windows executable files (.exe), which represent a subset of PE binaries. 

The MalwareBazaar repository is the primary source for the ransomware collection phase~\cite{malwarebazaar}. Besides, our dataset incorporates some ransomware samples from the dataset provided by Moreira et al.~\cite{moreira2023ransomware}. On the other hand, for the Benignware collection, samples were obtained again from by Moreira et al, the Benign-NET dataset, and the DikeDataset~\cite{moreira2023ransomware}, ~\cite{benignnet},~\cite{dikedataset}. 

To ensure architectural uniformity and compatibility with the analysis tools, only 32-bit PE (PE32) files were included in the dataset. Each sample was validated using the pefile Python library, which verified the file architecture and extracted relevant metadata for further classification tasks. Samples that did not conform to the expected format were excluded from the analysis pipeline.

Upon completion of sample collection, all collected samples underwent format and architecture validation to ensure that only 32-bit Windows PE files were retained. Additionally, all of the samples are scanned on the VirusTotal platform to ensure they are free from malware~\cite{virus}. 

To ensure comprehensive behavioral coverage, the dataset includes samples from 165 distinct ransomware families, spanning both prevalent strains—such as LockBit (134 samples), Conti (54 samples), Dharma (51 samples), Chaos (51 samples), and Phobos (43 samples)—and rare variants, including GpCode (1 sample), Erica (1 sample), and Crypute (1 sample).

The full distribution of these ransomware families is illustrated in Table~\ref{tab:ransomware_appendix} in the Appendix. This table presents each family name alongside its corresponding sample count, offering a clear overview of the dataset’s diversity and highlighting the class imbalance challenges considered during model development. As a result of these efforts, a total of 2,410 Windows 32-bit PE files are added to our collection, comprising 1,205 benignware samples and 1,205 ransomware samples.

\subsection{Static Feature Extraction and Analysis}

Static analysis involves the application of reverse engineering techniques to uncover the behavioral patterns of malware without executing the malicious code. This process is typically conducted within an isolated and secure environment to mitigate the risk of unintentional infections. Among the widely adopted tools in this domain, Radare2 serves as a prominent reverse engineering framework and is frequently utilized by malware researchers. 

In this study, we leveraged the Radare2 platform to statically analyze the collected ransomware and benignware executables. Additionally, its Python interface, r2pipe, enabled the seamless integration of reverse engineering capabilities into our automated analysis pipeline, allowing batch processing of multiple samples.

Each binary sample was loaded into the analysis environment using architecture-specific settings. The processor architecture was configured as 32-bit x86, and the system was set to operate in little-endian mode. To improve the accuracy of disassembly and maintain analytical consistency, several internal parameters were fine-tuned. These configurations ensured proper handling of relocations, efficient caching of binary data, and precise control over the analysis engine’s behavior. In particular, heuristic features—such as the skipping of no-operation (NOP) instructions or the inference of control flow beyond explicit instructions—were deliberately disabled to preserve deterministic and reproducible analysis outcomes. From the resulting disassembly output, opcode-level information was isolated by filtering instruction mnemonics against a predefined list of known opcodes. Only instructions matching relevant patterns were retained, and these were compiled into linear opcode sequences that served as static representations of each sample's execution logic.

As part of our static analysis procedure, we applied text mining techniques to extract relevant features. Initially, the OPCODEs from each sample are extracted and saved as individual text files. Subsequently, the frequency distribution of these OPCODEs was computed and stored in a structured format (.csv). Across the dataset, a total of 1002 unique OPCODEs were identified. The outcome of these steps was a structured dataset comprising 2410 records and 1002 features, which we refer to as OPCODE\_Dataset (OD). 

\subsection{Dynamic Feature Extraction and Analysis}

In order to extract runtime behavioral artifacts from Windows executables, we developed a modular dynamic analysis framework based on the Frida instrumentation toolkit. The system executes 32-bit Windows binaries within a controlled environment, injecting JavaScript-based probes to monitor runtime activity.

\begin{figure}[htbp]
  \centering
  \includegraphics[width=0.45\textwidth]{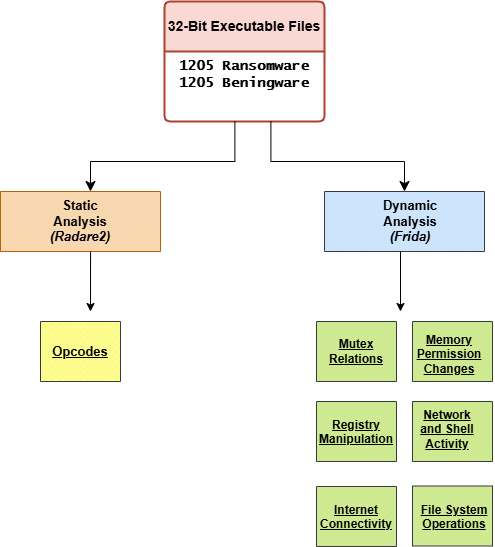}
  \caption{Overview of the hybrid feature extraction workflow combining static and dynamic analysis stages.}
  \label{fig:drawio78}
\end{figure}

Each sample is initially launched in a suspended state to allow for pre-execution instrumentation. Frida scripts are then injected to trace specific Windows API functions associated with registry access, mutex creation, file and network operations, and memory protection changes as seen in Figure \ref{fig:drawio78}. Intercepted API calls are transmitted to the host system and logged for subsequent analysis.

The entire framework is implemented in Python, which orchestrates sample execution, instrumentation, and termination under time-constrained conditions. Samples that have been previously analyzed are automatically skipped to ensure both efficiency and reproducibility. The modular design of the scripts supports both targeted and comprehensive monitoring, enabling scalable profiling of ransomware behaviors.

\begin{table*}[htbp]
\centering
\begin{scriptsize}
\caption{Monitored Behavioral Categories and Dynamic Features}
\label{tab:api_behavior_categories}
\begin{tabular}{|p{2.8cm}|p{5.1cm}|p{1.5cm}|p{6.8cm}|}
\hline
\textbf{Behavior Category} & \textbf{Instrumented Functions (Subset)} & \textbf{DLL Name} & \textbf{Purpose / Notes} \\
\hline
\textbf{Mutex Relation} & CreateMutex, CreateMutexEx (A/W) & \texttt{kernel32.dll} & Detects mutex creation for single-instance enforcement, thread synchronization, or sandbox evasion. \\
\hline
\textbf{Runtime Memory Permission Changes} & Page\_ReadOnly, Page\_ReadWrite, Page\_Execute, Page\_Execute\_Read & \texttt{kernel32.dll} & Monitors memory access right changes typically seen in unpacking or shellcode injection routines. \\
\hline
\textbf{Registry Manipulation} & RegCreateKey, RegSetValueEx, RegDeleteValue, RegOpenKey (A/W) & \texttt{advapi32.dll} & Tracks suspicious registry operations related to persistence, configuration changes, or evasion. \\
\hline
\textbf{Network and Shell Activity} & WSASend, ShellExecuteExW, GetAddrInfoExW & \texttt{ws2\_32.dll}, \texttt{shell32.dll} & Observes network transmissions, DNS resolution, and shell command execution. \\
\hline
\textbf{Internet Connectivity} & InternetOpenUrl, GetAddrInfo, GetAddrInfoEx & \texttt{wininet.dll}, \texttt{ws2\_32.dll} & Detects connection attempts to remote servers and possible C2 communications. \\
\hline
\textbf{File System Operations} & CreateFile, WriteFile, MoveFile, CopyFile, DeleteFile & \texttt{kernel32.dll} & Logs critical file actions including creation, overwrite, encryption, and deletion. \\
\hline
\end{tabular}
\end{scriptsize}
\end{table*}

\subsubsection{Mutex-Based Behavior Detection}

The framework includes instrumentation for the CreateMutex and CreateMutexEx functions (supporting both ANSI and Unicode variants), facilitating the detection of mutex-based techniques commonly employed by ransomware. These behaviors include enforcing single-instance execution, coordinating multi-threaded encryption routines, and identifying virtualized or sandboxed environments. Mutex names are extracted from memory and recorded in the behavioral logs.

\subsubsection{Runtime Memory Permission Changes}

Runtime hooks on the VirtualProtect function—exported from "kernel32.dll"—enable the detection of memory protection modifications. Such changes are typically used by ransomware to perform unpacking, execute shellcode, or evade static analysis. The memory protection flags are decoded into semantic labels (e.g., Page\_Execute\_ReadWrite) and recorded, allowing for the identification of just-in-time (JIT) code execution and memory-resident payload deployment.

\subsubsection{Registry Manipulation Monitoring}

Critical registry-related APIs such as "RegCreateKey", "RegOpenKey", "RegSetValueEx", and "RegDeleteValue" are hooked to monitor behaviors associated with persistence mechanisms, environment fingerprinting, and the deletion of forensic artifacts. These functions, exported from "Advapi32.dll", are instrumented in both their ANSI and Unicode variants. Registry handles (HKEYs) and associated key/value names are dynamically resolved to enable comprehensive behavioral logging.

\subsubsection{Network and Shell Activity Control}

The framework facilitates the interception and, if necessary, suppression of networking and shell execution API calls. Functions such as "WSASend", "GetAddrInfoExW", and "ShellExecuteExW" are instrumented to extract transmitted payloads, block domain name resolution, and control the execution of external commands. Runtime behavior is configurable through Frida’s messaging interface, allowing dynamic control over analysis policies.

\subsubsection{Internet Connectivity Monitoring}

To identify outbound communication attempts, hooks are applied to functions such as "InternetOpenUrl," "GetAddrInfo," and "GetAddrInfoEx." These functions reveal connections to command-and-control (C2) servers, suspicious domains, and staging infrastructures commonly leveraged in ransomware campaigns.

\subsubsection{File System Operations}

Low-level Win32 System API  calls such as "CreateFile," "WriteFile," "MoveFile," "CopyFile," and "DeleteFile" are monitored to track file system activities, including creation, modification, relocation, and deletion behaviors that are strongly indicative of ransomware encryption routines. During runtime, metadata such as file access permissions and creation modes is also extracted to support detailed behavioral profiling. These APIs are core components of the Windows API and are not associated with "wininet.dll" or "ws2\_32.dll"; prior references to these libraries in this context have been corrected accordingly.

This unified instrumentation system enables fine-grained tracing of ransomware behavior during execution, thereby supporting both real-time detection and post-infection forensic attribution. As a result of these efforts, a structured dataset comprising 2411 records and 5024 features, which we refer to as API\_Call\_Dataset (ACD).

Table~\ref{tab:api_behavior_categories} shows the mapping between observed DLLs and API functions included in our dynamic feature set. This mapping facilitates modular extensibility and compatibility with other behavioral datasets. It is also important to focus on behavioral signals that can be observed in the first seconds of execution to enable early detection during dynamic analysis.

\subsection{Feature Integration and Model Training}

In this study, a hybrid dataset —RansomTrack— was constructed by combining the outputs of both static and dynamic analysis techniques. This integration aims to leverage the complementary strengths of these two approaches to develop an effective detection mechanism against ransomware.

To identify features with higher discriminative power within each class, the dataset was partitioned into static and dynamic components. Each feature set was further subdivided based on binary class labels (label = 1 for ransomware, label = 0 for benignware), and Pearson correlation analysis was independently applied within each class.

Based on the resulting intra-class correlation patterns, features were categorized into three primary groups:

\begin{itemize}
    \item \textbf{Class-Neutral Highly Correlated Features (correlation $\geq$ 0.9 in both classes):} These features form strongly correlated clusters consistently observed across both ransomware and benignware samples. Their consistent presence suggests that they capture structural or behavioral properties common to all executable files.
    \item \textbf{Benignware-Specific Correlated Features (correlation $\geq$ 0.7 for label = 0.0 and $\leq$ 0.0 for label = 1.0):} These features exhibit moderate to strong correlations within benignware samples but show weak or negative correlation in ransomware. This contrast indicates features that are more representative of benign behavior.
    \item \textbf{Ransomware-Specific Correlated Features (correlation $\geq$ 0.7 for label = 1.0 and $\leq$ 0.0 for label = 0.0):} Conversely, these features are strongly correlated within ransomware samples but show weak or inverse relationships in benignware. Such features may reflect malicious or evasive behaviors that are characteristic of ransomware.
    
\end{itemize}

\begin{figure}[htbp]
  \centering
  \includegraphics[width=0.45\textwidth]{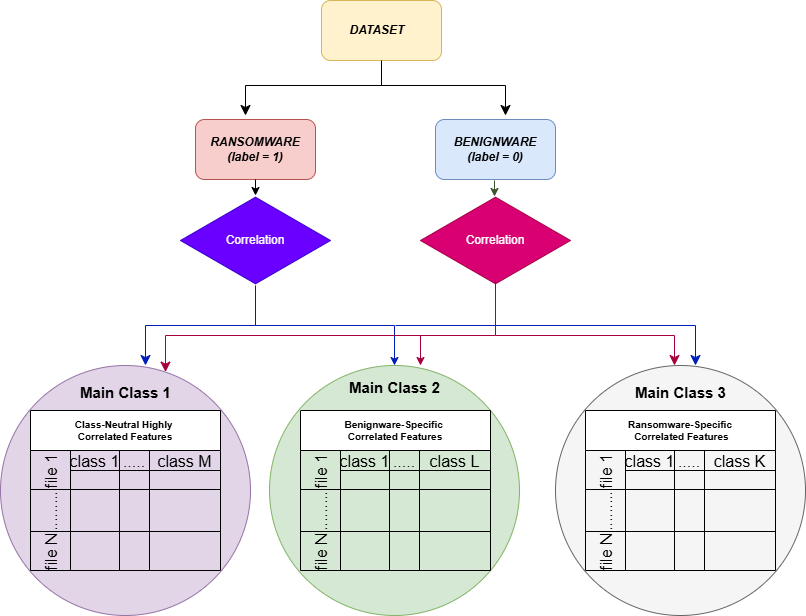}
\caption{Feature Integration Overview}
  \label{fig:drawi}
\end{figure}

This correlation-based categorization provides deeper insight into class-specific feature dependencies and enhances the interpretability of both static and dynamic attributes in the dataset, as illustrated in Figure ~\ref{fig:drawi}.

The threshold values used for feature grouping follow conventional interpretations of Pearson correlation strength: coefficients between 0.7 and 0.9 are considered moderately correlated, values above 0.9 indicate strong correlation, and values equal to or below 0.0 are regarded as uncorrelated or negatively correlated~\cite{calkins_correlation}.

To evaluate the effectiveness of the extracted features, a comprehensive set of classification algorithms was employed. The dataset was partitioned into training and testing subsets using an 80/20 stratified split to maintain class balance. A fixed random seed was applied to ensure the reproducibility of experimental results.

The classification models used in this study are organized into five main categories:

\begin{itemize}
    \item \textbf{Tree-based classifiers:}Decision Tree, Random Forest, and Extremely Randomized Trees, which are well-suited for modeling nonlinear feature interactions.
    \item \textbf{Boosting algorithms:}XGBoost, Gradient Boosting, AdaBoost, and CatBoost, known for their ability to iteratively improve weak learners and enhance predictive accuracy.
    \item \textbf{Neural network model:}A Multi-Layer Perceptron (MLP) with multiple hidden layers, capable of capturing complex, high-dimensional feature representations.
    \item \textbf{Linear model:}Logistic Regression, employed to assess baseline separability within the feature space.
    \item \textbf{Distance and kernel based methods:}K-Nearest Neighbors and Support Vector Machine, which rely on similarity metrics and kernel-defined decision boundaries.
\end{itemize}

To enhance classification robustness and reduce individual model bias, a soft voting ensemble strategy was adopted, aggregating predictions from multiple base classifiers. Hyperparameter tuning was performed for selected models—such as Random Forest—using grid search combined with five-fold cross-validation to optimize performance. All models were implemented using widely adopted machine learning libraries, ensuring consistency, scalability, and reproducibility across all experimental settings.

\section{Experimental Results} 
\label{sec:results}

To evaluate the effectiveness of the proposed hybrid framework, we conducted experiments on the RansomTrack dataset, which contains 165 ransomware families and 2,410 labeled instances in their corresponding benign binaries.

All phases of the proposed framework—including static and dynamic analysis, feature extraction, data preprocessing, machine learning model training, and testing—were conducted on a virtual machine configured using Oracle VirtualBox. The virtual environment runs a Windows 10 Pro 64-bit operating system with 8 GB of RAM and 2 virtual CPUs.  

The evaluation of the proposed framework is carried out from multiple perspectives, including malware analysis and feature extraction, data preprocessing and dimensionality reduction, ransomware detection, model interpretability using SHAP (SHapley Additive exPlanations), model-specific false negative analysis, and runtime performance analysis.

\subsection{Malware Analysis and Feature Extraction}

\begin{figure*}[ht]
  \centering
  \includegraphics[scale=0.45]{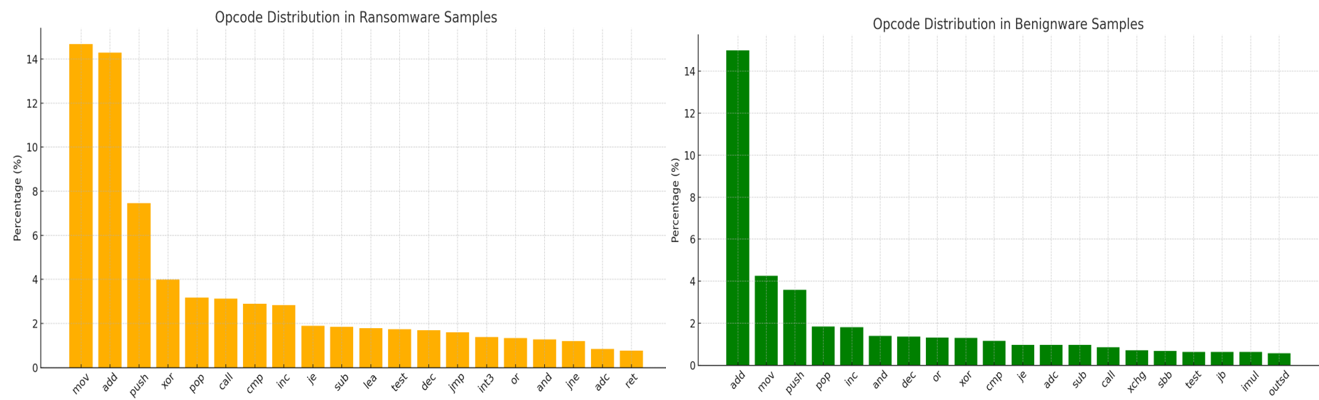}
  \caption{Opcode distributions in ransomware and benignware samples}
  \label{fig:opcode-distibution}
\end{figure*}

First of all, OPCODE distributions are compared between ransomware and benignware samples. In ransomware samples, the "MOV" instruction constitutes approximately 14.5\% of all opcodes, highlighting its central role in memory-related operations such as key loading, buffer manipulation, and data transfer during encryption processes. Other frequently observed instructions include "ADD, XOR, CALL, and INT3." The distribution of OPCODE usage among ransomware samples is depicted in Figure~\ref{fig:opcode-distibution}. Of particular note, the INT3 instruction is often utilized for anti-debugging purposes, enabling malware to detect the presence of debugging or sandboxing environments. The XOR instruction, by contrast, is commonly employed for obfuscation and lightweight encryption, especially in self-decrypting or packed code segments. Overall, ransomware binaries exhibit a strong dependency on control-flow and memory-manipulation instructions, which aligns with typical behaviors such as payload unpacking, evasion, and runtime code modification.

In contrast, for the benignware samples as shown in Figure~\ref{fig:opcode-distibution}, instructions such as "MOV" and "ADD" remain prevalent, the overall distribution is more uniform, and it lacks a dominant presence of opcodes associated with anti-analysis or encryption-related activities. Instructions like "PUSH", "POP", and various arithmetic operations appear at moderate frequencies, indicative of standard application logic and typical control-flow behavior rather than malicious functionality.

\begin{figure*}[ht]
  \centering
  \includegraphics[scale=0.21]{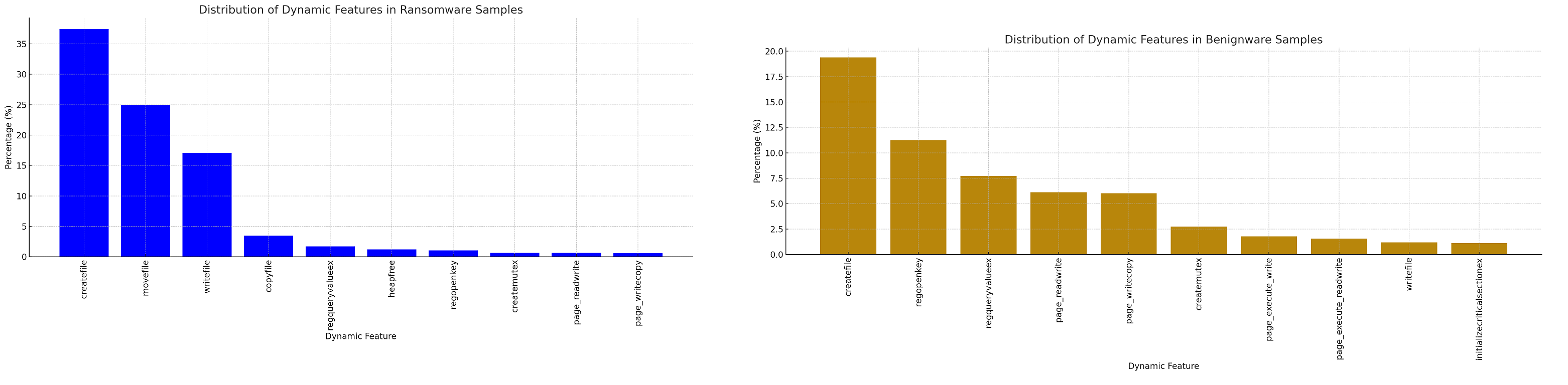}
  \caption{Dynamic feature distribution in ransomware and benignware samples}
  \label{fig:dynamic-distibution}
\end{figure*}

Our second findings are obtained from dynamic analysis. The most frequently invoked API calls by ransomware samples include "CreateFile," "MoveFile," and "WriteFile," all of which are central to ransomware's core functionality—namely, file encryption, renaming, and unauthorized data manipulation. Additionally, registry access routines such as "RegQueryValueEx" and memory protection changes like "Page\_Execute\_ReadWrite" and "Page\_ReadWrite" are observed with high frequency, suggesting the presence of behaviors such as shellcode injection, unpacking, and in-memory payload execution. Figure~\ref{fig:dynamic-distibution} presents the distribution of dynamic features extracted from ransomware and beningware samples.

Less frequent API calls, such as "DeleteFile," "EncodePointer," and "FlsAlloc," may be associated with anti-forensic actions, obfuscation, or resource allocation mechanisms that support stealth and persistence.

In contrast, for benignware samples, registry-related operations such as "RegOpenKey" and "RegQueryValueEx" are dominant, along with standard file access via "CreateFile," indicative of routine system configuration and application initialization. Notably, benign samples exhibit minimal manipulation of memory permissions and lack aggressive file-handling behavior, which distinguishes them from ransomware.

Furthermore, the presence of API calls like "EvtFormatMessage," "OpenProcessToken," and "GetTokenInformation" suggests legitimate security context queries and event logging functionality, commonly found in administrative or utility software.

These visualized distributions across Figures~\ref{fig:opcode-distibution} and \ref{fig:dynamic-distibution} reinforce the behavioral divergence between ransomware and benignware, both in static opcode patterns and dynamic API usage. This contrast supports the design of interpretable and discriminative hybrid feature models for ransomware detection.

\subsection{Data Preprocessing and Dimensionality Reduction}

We applied correlation-based feature classification to the static opcode features extracted from disassembled PE32 samples. By analyzing co-occurrence patterns and semantic relationships among opcodes, we identified meaningful clusters associated with common programming constructs, control-flow mechanisms, and obfuscation techniques.

The resulting opcode groups exhibit class-specific tendencies—some are predominantly observed in ransomware samples (e.g., obfuscation-related opcodes such as xor, nop, and int3), while others are more frequent in benignware (e.g., control-transfer instructions like ret, call, or data movement instructions used in initialization routines). Shared opcode clusters typically represent foundational system-level instructions present in both malware and legitimate software.

This grouping approach not only improves interpretability during model training but also contributes to dimensionality reduction without compromising discriminative power. Representative static opcode clusters and their associated behavioral interpretations are summarized in Table~\ref{tab:feature-groups}.

\begin{table*}[ht]
\centering
\begin{scriptsize}
\caption{Static and Dynamic Feature Groups via Correlation-Based Feature Clustering}
\label{tab:feature-groups}
\begin{tabular}{|p{2.0cm}|p{2.9cm}|p{4cm}|p{3cm}|p{4cm}|}
\hline
\textbf{Group Type}                         & \textbf{Static Feature Examples}                   & \textbf{Static Feature Description}                                                                                & \textbf{Dynamic Feature Examples}                                                           & \textbf{Dynamic Feature Description}                                                                 \\ \hline
\multirow{2}{*}{\shortstack{Benignware\\Specific Groups}} & \textit{bndmov, vmsave, vmxoff, skinit}                     & Virtualization and system state management; commonly found in clean system exits or virtualization-aware software. & \textit{adjustwindowrect, drawtexta, fillrect, getcursor, setfocus}                                   & Graphical UI rendering and user interaction management; typical of GUI-based legitimate applications. \\ \cline{2-5} 
                                            & \textit{vpmacsdql, vpsllvq, vpblendd}                       & Vector arithmetic operations used in multimedia, graphics, and scientific computations.                            & \textit{getdateformatw, gettimeformatw, getlocaleinfoa, gettimezoneinformation}                       & Localization, formatting, and time/date retrieval routines used in user-facing programs.              \\ \cline{2-5} 
                                            & \textit{bsr, movdqu, pxor, pshufb}                          & Legacy I/O and memory operations for structured buffer handling in user-level applications.                        & \textit{regsetvalueexw, regdeletevaluew, setenvironmentvariablew}                                     & Non-destructive registry access or benign configuration settings by installers and applications.      \\ \cline{2-5} 
                                            & \textit{cvttpd2dq, vcvttsd2si, vcomisd}                     & Data conversion and comparison instructions frequently observed in benign numerical processing.                    & \textit{createfile, mapviewoffile, heapcompact, heapfree}                                             & Safe file and memory operations—typical of setup tools, editors, or document-handling utilities.      \\ \hline
\multirow{2}{*}{\shortstack{Ransomware \\ Specific Groups}} & \textit{vpminsd, vpor, vpaddd, sha1msg1, sha256msg2}        & Cryptographic and vectorized hashing operations typical in ransomware encryption routines.                         & \textit{copyfilew, createfilemappinga, movetoex, ntcreatefile}                                        & File locking, replication, and direct mapping—often used in encryption and destructive overwrite.     \\ \cline{2-5} 
                                            & \textit{movntdq, vmovntdq, pshufb, movhlps}                 & Streaming memory writes used for fast data overwrite and stealth payload delivery.                                 & \textit{regdeletekeya, regsetvalueexa, regenumkeyexa, regqueryinfokeya}                               & Registry manipulation patterns used for persistence, disabling recovery, or altering boot behavior.   \\ \cline{2-5} 
                                            & \textit{xend, mwaitx, xsetbv}                               & Low-level control instructions related to system context switching and VM escape techniques.                       & \textit{createprocessw, getcurrentthread, getcurrentthreadid, terminatethread}                        & Process/thread hijacking or spawning subprocesses; used to evade or disable protection mechanisms.    \\ \cline{2-5} 
                                            & \textit{int3, call, cmpneqps, sfence}                       & Control flow manipulation and timing-based obfuscation techniques for anti-debugging or evasive execution.         & \textit{setdlldirectoryw, shellexecuteexa, loadlibrarya}                                              & DLL injection, side-loading, or stealth execution routines supporting payload deployment.             \\ \hline
\multirow{2}{*}{\shortstack{Shared Feature\\ Groups}}      & \textit{fadd, fmul, fcom, fsqrt, fucompp}                   & Floating-point arithmetic common to both benign applications and malicious cryptographic routines.                 & \textit{strcpy, strcat, \_itoa, \_stricmp, wcsncmp}                                                   & Standard string and buffer routines shared across most applications, benign or malicious.             \\ \cline{2-5} 
                                            & \textit{int, hlt, iret, cli, sti, pushf}                    & System-level and interrupt management instructions found in most executable flows.                                 & \textit{getmodulehandlew, getstartupinfow, loadimagew}                                                & Process initialization and dynamic linking support used during startup in various software.           \\ \cline{2-5} 
                                            & \textit{mov, push, pop, lods, stos, xchg}                   & Basic data movement and stack operations essential for any control-flow logic.                                     & \textit{regopenkeyexa, regqueryvalueexa, openscmanagerw, queryservicestatus}                        & Registry/service interaction routines used by installers, utilities, or malware loaders alike.        \\ \cline{2-5} 
                                            & \textit{vaddps, vcvtsd2si, vminps, vpaddb, vpsubd, vpsrldq} & Vector arithmetic patterns observed both in ransomware and benign multimedia software.                             & \textit{cryptacquirecontexta, cryptgenrandom, convertstringsecurity- descriptortosecuritydescriptorw} & Cryptographic context setup—can appear in secure apps and ransomware key handling.                    \\ \hline
\end{tabular}
\end{scriptsize}
\end{table*}

Similarly, dynamic behavioral features extracted from API call traces and memory activity logs were grouped using intra-class Pearson correlation analysis. Unlike conventional high-level API call profiling, our method captures low-level runtime artifacts for instance "Memory protection changes," "Heap allocations," "Registry key manipulations," "File system operations," and "Network and shell invocation" routines.

This analysis revealed distinct behavior profiles for ransomware and benignware. For example, ransomware-specific clusters include patterns of memory-resident payload execution, mutex-based evasion, and aggressive file renaming or deletion. In contrast, benignware groups were characterized by registry initialization routines, event log accesses, and standard file read/write behavior.

All identified feature groups are categorized as benignware-specific, ransomware-specific, or shared in Table~\ref{tab:feature-groups}, providing a clear behavioral taxonomy.

\begin{table*}[ht]
\centering
\begin{scriptsize}
\caption{Hybrid Feature Clustering}
\label{tab:feature-clusters}
\begin{tabular}{|p{2.0cm}|p{4cm}|p{4cm}|p{6cm}|}
\hline
\textbf{Group Type}                         & \textbf{Feature Group}                          & \textbf{Representative Features}                                                     & \textbf{Description}                                                                      \\ \hline
\multirow{2}{*}{\shortstack{Benignware\\Specific Groups}} & GUI Rendering and Input Handling                & \textit{gdipcreatebitmapfromhbitmap, loadbitmapw, dispatchmessagew, getmessagew}     & Common in interactive GUI applications; reflects standard rendering pipelines.            \\ \cline{2-4} 
                                            & Registry and DLL Handling                       & \textit{cryptacquirecontextw, getfileversioninfoa, regqueryinfokeya, loadlibraryexw} & Used for querying system version info and handling optional libraries in benign programs. \\ \cline{2-4} 
                                            & Memory Mapping and Process Utilities            & \textit{createfilemappinga, mapviewoffile, duplicatehandle, getthreaddesktop}        & Typical of system configuration tools, benign services, or sandbox utilities.             \\ \cline{2-4} 
                                            & Numeric and Vector Operations (Static)          & \textit{cvtdq2pd, vcvtpd2dq, vpmuludq, movhpd}                                       & Indicates benign computational routines involving multimedia or scientific tasks.         \\ \cline{2-4} 
                                            & String and Localization Utilities               & \textit{comparestringa, getlocaleinfoa, gettimezoneinformation}                      & Reflects internationalization and regional format handling in user applications.          \\ \hline
\multirow{2}{*}{\shortstack{Ransomware \\ Specific Groups}} & Cryptographic and Vector Payloads (Static)      & \textit{vpminsd, vfmsub213sd, sha1msg1, sha256msg2, kandnb}                          & Corresponds to cryptographic operations, likely encryption core logic.                    \\ \cline{2-4} 
                                            & Persistence and Obfuscation APIs                & \textit{createprocessw, loadlibrarya, setsecuritydescriptorowner, vmovapd, vmovups}  & Used to hide ransomware payloads or maintain access after reboot.                         \\ \cline{2-4} 
                                            & System and Thread Control                       & \textit{getlocaleinfow, getstdhandle, getwindowtextlengtha, timegettime}             & Includes timing and system enumeration calls used in sandbox detection or evasion.        \\ \cline{2-4} 
                                            & Destructive or Privileged Instructions (Static) & \textit{smsw, vsqrtss, mwaitx, vmrun, lldt}                                          & Low-level processor instructions linked to control privilege levels or system state.      \\ \cline{2-4} 
                                            & Resource and Dialog Manipulation                & \textit{createpen, drawthemetextex, screentoclient}                                  & May be used to mimic or obstruct legitimate user interface elements.                      \\ \hline
\multirow{2}{*}{\shortstack{Shared Feature\\ Groups}}      & Common Windows APIs                             & \textit{createwindowexa, getmessagea, loadcursora, postmessagea}                     & Fundamental API calls required for basic GUI event handling.                              \\ \cline{2-4} 
                                            & String and Locale Utilities (Static + Dynamic)  & \textit{cvtsi2sd, movss, xorps, wcsstr, lcmapstringw}                                & Used in string formatting and parsing routines across both benign and malicious software. \\ \cline{2-4} 
                                            & Process and Memory Control                      & \textit{getcurrentprocess, tlsalloc, sleepconditionvariablecs, pmuludq}              & General resource handling features present in various PE workflows.                       \\ \cline{2-4} 
                                            & File and Path Utilities                         & \textit{findfirstfilea, pathcombinew, getvolumeinformationw}                         & Involved in accessing system directories and file enumeration.                            \\ \cline{2-4} 
                                            & Security and Registry APIs                      & \textit{cryptacquirecontexta, setfilesecurityw, regenumkeya, deleteobject}           & Shared between installer utilities and malware routines.                                  \\ \hline
\end{tabular}
\end{scriptsize}
\end{table*}

To explore the interplay between static and dynamic analysis, a hybrid correlation analysis is conducted by combining both feature domains. Static features included opcode distributions, while dynamic features encompassed behavioral indicators such as memory modifications and API sequences.

Interestingly, the resulting clusters showed that no hybrid group contained a mix of static and dynamic features, highlighting a strong orthogonality between the structural and behavioral domains. This finding suggests that ransomware exhibits distinct operational layers: static indicators related to packing and obfuscation, and dynamic traits related to malicious execution logic.

The absence of cross-domain co-correlation emphasizes the value of hybrid analysis—each domain captures unique and complementary signals for robust ransomware detection. The representative feature clusters resulting from the hybrid correlation analysis are presented in Table~\ref{tab:feature-clusters}, corresponding to benignware-specific, ransomware-specific, and class-shared behavioral patterns. While beingware typically emphasizes user interface manipulation, registry querying, memory mapping, and internationalization, ransomware uses more aggressive and sophisticated features such as encryption algorithms, obfuscation methods, system control, and user interface emulation. There are some common sets of APIs shared by both types, such as basic Windows interface calls, string manipulation functions, and helper functions related to file paths, which are used in both benign and malicious software. This suggests that malware detection should focus not only on individual API calls, but also on their combinations and behavioral patterns.

\subsection{Ransomware Detection}

\begin{table*}[ht]
\centering
\caption{Per-Sample Training and Testing Times for Random Forest and Logistic Regression}
\label{tab:full_rf_lr_comparison}
\renewcommand{\arraystretch}{1.20}
\begin{scriptsize}
\begin{tabular}{|l|l|c|c|c|c|c|c|c|c|c|c|c|}
\hline
\textbf{Model} & \textbf{Feature} & \textbf{Acc.} & \textbf{Prec.} & \textbf{Rec.} & \textbf{Spec.} & \textbf{G-Mean} & \textbf{F1} & \textbf{ROC-AUC} & \textbf{Bal. Acc.} & \textbf{Log Loss} & \textbf{\shortstack {Train Time \\ Per Sample (s)}} & \textbf{\shortstack{Test Time \\ Per Sample (s)}} \\
\hline
RF & Static  & 0.91 & 0.96 & 0.85 & 0.97 & 0.90 & 0.90 & 0.98 & 0.91 & 0.21 & 0.0008 & 0.0001 \\
RF & Dynamic & 0.94 & 0.91 & 0.97 & 0.91 & 0.94 & 0.94 & 0.98 & 0.94 & 0.19 & 0.0012 & 0.0002 \\
RF & Hybrid  & \textbf{0.95} & \textbf{0.95} & \textbf{0.95} & \textbf{0.95} & \textbf{0.95} & \textbf{0.95} & \textbf{0.99} & \textbf{0.95} & \textbf{0.14} & \textbf{0.0016} & \textbf{0.0002} \\
\hline
LR & Static  & 0.82 & 0.95 & 0.68 & 0.97 & 0.81 & 0.79 & 0.90 & 0.82 & 0.56 & 0.0061 & 0.000001 \\
LR & Dynamic & 0.94 & 0.93 & 0.95 & 0.93 & 0.94 & 0.94 & 0.98 & 0.94 & 0.17 & 0.0062 & 0.00001 \\
LR & Hybrid  & 0.93 & 0.93 & 0.93 & 0.93 & 0.93 & 0.93 & 0.98 & 0.93 & 0.22 & 0.0132 & 0.00001 \\
\hline
\end{tabular}
\end{scriptsize}
\end{table*}

\begin{table*}[ht]
\centering
\caption{Evaluation Metrics of Classifiers on Hybrid Feature Set with Per-Sample Timing}
\label{tab:hybrid_classifiers}
\renewcommand{\arraystretch}{1.15}
\scriptsize
\begin{tabular}{|l|c|c|c|c|c|c|c|c|c|c|c|}
\hline
\textbf{Model} & \textbf{Acc.} & \textbf{Pre.} & \textbf{Rec.} & \textbf{Spec.} & \textbf{G-Mean} & \textbf{F1} & \textbf{ROC-AUC} & \textbf{Bal. Acc.} & \textbf{Log Loss} & \textbf{\shortstack {Train Time \\ Per Sample (s)}} & \textbf{\shortstack{Test Time \\ Per Sample (s)}} \\
\hline
RF           & 0.95 & 0.95 & 0.95 & 0.95 & 0.95 & 0.95 & 0.99 & 0.95 & 0.14 & \textbf{0.0016} & \textbf{0.0002} \\
LR           & 0.93 & 0.93 & 0.93 & 0.93 & 0.93 & 0.93 & 0.98 & 0.93 & 0.22 & 0.0132 & 0.000001 \\
KNN          & 0.95 & 0.97 & 0.93 & 0.98 & 0.95 & 0.95 & 0.97 & 0.95 & 0.81 & 0.0217 & 0.0013 \\
DT           & 0.94 & 0.95 & 0.93 & 0.95 & 0.94 & 0.94 & 0.96 & 0.94 & 1.34 & \textbf{0.0008} & \textbf{0.00001} \\
MLP          & 0.94 & 0.92 & 0.96 & 0.92 & 0.94 & 0.94 & 0.98 & 0.94 & 0.16 & 0.1251 & 0.0001 \\
XGBoost      & \textbf{0.96} & \textbf{0.97} & 0.95 & \textbf{0.98} & 0.96 & 0.96 & \textbf{0.99} & 0.96 & \textbf{0.12} & 0.0025 & 0.0001 \\
Soft Voting  & \textbf{0.96} & \textbf{0.97} & \textbf{0.96} & 0.97 & \textbf{0.96} & \textbf{0.96} & \textbf{0.99} & \textbf{0.96} & 0.16 & 0.1273 & 0.0091 \\
\hline
\end{tabular}
\end{table*}

\begin{figure*}[htbp]
    \centering
    \includegraphics[scale=0.40]{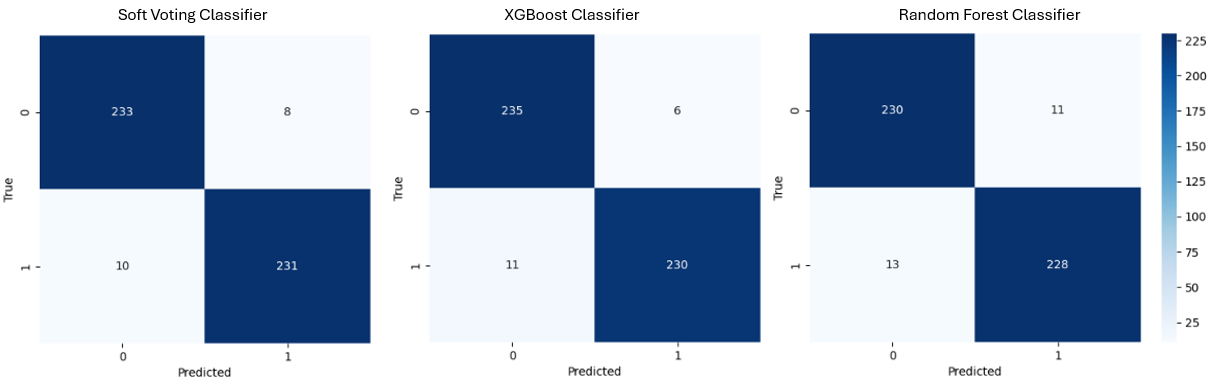}

\caption{Confusion matrices}
\label{fig:conf_matrices}
\end{figure*}

Our previous research reveals that Random Forest (RF) outperforms classical machine learning algorithms such as Decision Tree (DT) and XGBoost in ransomware detection, particularly when leveraging dynamic behavioral features like API call sequences~\cite{caliskan}. Similarly, Yüksel~\cite{yuksel2025} reports that RF achieves superior accuracy on API-based datasets due to its ability to capture nonlinear patterns and handle high-dimensional feature spaces effectively.

Building on these insights, we prioritized RF for its demonstrated strength in modeling behavioral data. For comparison and baseline interpretability, we also included logistic regression (LR), which is known for its effectiveness in linearly separable feature spaces. Prior work by Mowri et al.~\cite{mowri2022} supports LR’s performance in such settings, particularly when applied to subsets of static features.

To systematically evaluate the contrast between linear and nonlinear classifiers, we independently trained RF and LR on three distinct feature representations: First, static features derived from opcode distributions, then dynamic features extracted from API call traces, and lastly, a hybrid set combining both. This structured approach allowed us to assess how model performance shifts as the feature space transitions from structural to behavioral representations and their integration.

Following this RF–LR comparison, we extended our evaluation to include additional classifiers—namely DT, K-Nearest Neighbors (KNN), Multi-Layer Perceptron (MLP), and XGBoost. To further improve classification robustness and stability, we implemented a Soft Voting Ensemble model that aggregates the predictions from these diverse models using a weighted consensus mechanism.

In Tables~\ref{tab:full_rf_lr_comparison} and~\ref{tab:hybrid_classifiers}, 
The performance results of all classifiers are reported. Table~\ref{tab:full_rf_lr_comparison} presents a detailed comparison of RF and LR across all three feature representations using multiple metrics, including accuracy, precision, recall, specificity, G-Mean, F1-score, ROC-AUC, balanced accuracy, log loss, and runtime. The results highlight clear performance gaps between Random Forest and Logistic Regression across different feature sets. Random Forest (RF) demonstrates consistent superiority, particularly on the hybrid feature configuration, achieving 0.95 accuracy, 0.95 F1-score, and a near-perfect 0.99 ROC-AUC, while maintaining a low log loss of 0.14. In contrast, Logistic Regression (LR) underperforms on static features (recall: 0.68, F1: 0.79), reflecting its limitations in modeling nonlinear decision boundaries.

Table~\ref{tab:hybrid_classifiers} focuses exclusively on the hybrid representation and compares all classifiers to assess their generalization capability over this enriched feature space. In this comparison,  a broader set of classifiers was evaluated on the hybrid feature set. Among them, ensemble-based models—specifically XGBoost and Soft Voting—achieved the highest overall performance, with both reaching 0.96 accuracy, 0.97 precision, 0.96 F1-score, and 0.99 ROC-AUC. However, closer inspection of their confusion matrices reveals meaningful distinctions in misclassification behavior and runtime characteristics.

To complement the tabular evaluations and provide visual insights into classification behavior, we generated confusion matrices for three key models—Soft Voting, XGBoost, and RF—on the held-out 20\% test set. These matrices reveal how well each model distinguishes between benignware (class 0) and ransomware (class 1).

Figure~\ref{fig:conf_matrices} shows the confusion matrices for the Soft Voting, XGBoost, and RF Classifiers. The Soft Voting achieved the best overall performance: 231 true positives with only 10 false negatives, indicating high sensitivity. Additionally, it produced 233 true negatives and just 8 false positives, reflecting a strong specificity. XGBoost's confusion matrix demonstrates comparable performance, with 230 true positives and 11 false negatives. Notably, it yielded only 6 false positives, suggesting a slightly more conservative decision boundary. RF's results are consistent with the other top models but with a marginally higher false negative count (13), indicating slightly lower sensitivity to class 1 instances. This makes Soft Voting the most sensitive model to ransomware detection. However, it comes at a substantial computational cost: its training and inference times were measured at 245.35s and 4.37s, respectively, whereas XGBoost completed the same tasks in just 4.82s and 0.05s. Random Forest required moderate time (8.76s training, 0.24s inference).

Overall, the confusion matrices reinforce the quantitative findings and highlight the benefits of ensemble-based methods, particularly in minimizing false negatives, which is a critical objective in ransomware detection systems.

\subsection{Interpreting Model Predictions with SHAP}

\begin{figure*}[ht]
    \centering
    \includegraphics[scale=0.40]{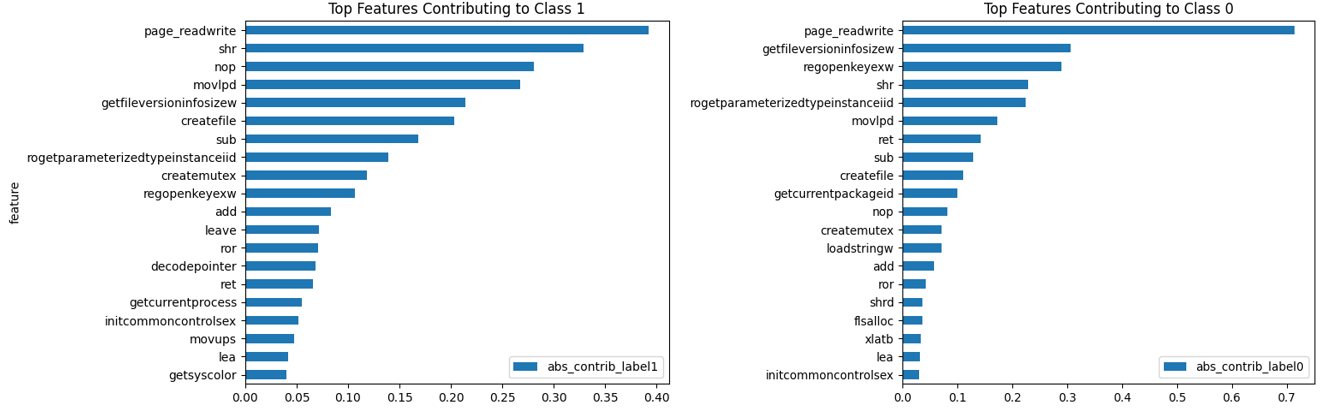}
    \caption{Top 20 features contributing to class 1 (top) and class 0 (bottom) based on absolute mean SHAP values.}
    \label{fig:shap_contrib_classwise}
\end{figure*}

To enhance model interpretability, we used SHAP (SHapley Additive exPlanations) values to analyze the feature contributions of the trained XGBoost classifier. SHAP assigns an importance score to each feature based on its marginal contribution to individual predictions.

We computed average SHAP values separately for class 0 and class 1 samples to identify the most influential features for each class. Features were ranked based on the absolute mean SHAP values and visualized in class-wise bar plots, offering a detailed view of decision rationale for both benign and ransomware classifications.

The SHAP-based analysis in Figure ~\ref{fig:shap_contrib_classwise} reveals distinct behavioral and structural patterns characteristic of ransomware and highlights features commonly associated with benign application behavior.

In ransomware behavior, memory-related operations such as \verb|page_readwrite|, \verb|getcurrentprocess|, and \verb|decodepointer| indicate active runtime memory manipulation---a common trait in unpacking, code injection, and self-modifying malware. A substantial portion of highly influential features consists of low-level assembly instructions including \texttt{nop}, \texttt{shr}, \texttt{movlpd}, \texttt{add}, \texttt{sub}, and \texttt{ror}, which are f and \texttt{regopenkeyexw} reflect direct interaction with file systems and the Windows Registry, reinforcing patterns like mass file access, sandbox detection, and persistence mechanisms. Collectively, these results indicate that the model effectively learns operational signatures specific to ransomware across memory, instruction-level, and API-based behavioral domains.

In benignware behavior, although certain operations such as \verb|page_readwrite| and \verb|regopenkeyexw| are shared with ransomware, their usage in benign samples reflects legitimate system interactions—such as configuration loading, structured file access, or version metadata retrieval—rather than malicious intent. Low-level instructions like \texttt{shr}, \texttt{movlpd}, \texttt{ret}, \texttt{sub}, \texttt{add}, and \texttt{ror} also appear in benignware, but typically in contexts generated by standard compilers, rather than obfuscated loops or control-flow tampering. Furthermore, modern API calls such as \texttt{rogetparameterizedtypeinstanceiid}, \texttt{getcurrentpackageid} and \texttt{loadstringw} are frequently observed in benign samples, indicating structured user interface initialization, resource localization, and system-level querying, particularly within UWP or WinRT-based applications.

In contrast, a detailed SHAP-based comparison identifies discriminative instructions and API calls that are uniquely associated with ransomware behavior. Instructions such as \texttt{nop}, \texttt{leave}, and \texttt{movups} are observed exclusively in ransomware samples and are widely recognized as indicative of evasive behavior. These instructions are commonly employed in obfuscation routines, return-oriented programming (ROP) chains, or shellcode deployment, suggesting deliberate attempts to hinder static and dynamic analysis by security tools.

Distinctive assembly-level features also exist on the benign side. Instructions such as \texttt{shrd} and \texttt{xlatb} appear only in benign samples and are often the result of performance-optimized code generation or low-level compilation artifacts. These patterns do not reflect any suspicious behavior and align with expected software execution flows.

At the behavioral level, several API calls show class-specific usage patterns. Exclusively in ransomware samples, APIs such as \texttt{getcurrentprocess}, \texttt{decodepointer}, and \texttt{getsyscolor} reflect anti-analysis strategies, memory handle manipulation, and runtime environment discovery. These calls enable the malware to gather context information about its execution environment and adapt accordingly to evade detection.

Conversely, benign-exclusive API usage, including \texttt{get\allowbreak current\allowbreak package\allowbreak id}, \texttt{loadstringw}, and \texttt{fsalloc}, points to legitimate application functionality such as localized GUI rendering, application metadata querying, and file I/O initialization.

Together, these observations confirm the model’s ability to differentiate not only based on feature presence, but also through an understanding of their contextual and semantic roles within executable workflows. Such nuanced insights significantly enhance the robustness and interpretability of ransomware detection models.

\subsection{Model-Specific False Negative Analysis}

To quantify how frequently each model misclassifies ransomware samples (class 1), we computed the false negative rate (FNR), defined in Equation~\ref{eq:fnr}:

\begin{equation}
\text{FNR (\%)} = \frac{\text{FN}}{\text{TP} + \text{FN}} \times 100
\label{eq:fnr}
\end{equation}

\noindent where FN denotes the number of false negatives and TP denotes the number of true positives. A higher FNR indicates a model's failure to detect ransomware, which is especially critical in cybersecurity applications.

The computed FNR values for the top-performing classifiers are summarized in Table~\ref{tab:fnr-results}.

\begin{table}[ht]
\centering
\caption{False Negative Rates (FNR) for Ransomware Samples}
\label{tab:fnr-results}
\begin{tabular}{|l|c|c|}
\hline
\textbf{Model} & \textbf{FN / (TP + FN)} & \textbf{FNR (\%)} \\
\hline
Soft Voting     & $10 / 241$ & \textbf{4.15\%} \\
XGBoost         & $11 / 241$ & \textbf{4.56\%} \\
Random Forest   & $13 / 241$ & \textbf{5.39\%} \\
\hline
\end{tabular}
\end{table}

Frequently observed in custom encryption routines and obfuscation logic. Furthermore, API calls such as \texttt{createfile}, \texttt{createmutex},These results emphasize that while all three models—Soft Voting, XGBoost, and Random Forest—are highly effective on hybrid features, their optimal usage depends on deployment context. Soft Voting achieves the highest detection coverage with minimal ransomware misclassification, making it ideal for high-assurance scenarios such as critical infrastructure or forensic analysis, where recall is paramount and runtime constraints are secondary. Its ensemble architecture enhances robustness by capturing diverse decision boundaries, improving performance against ambiguous or borderline samples.

In contrast, XGBoost delivers near-instant inference (0.05s) with slightly higher misclassification rates, making it well-suited for low-latency environments such as endpoint protection or online threat detection systems. Random Forest strikes a balance between recall and runtime, offering reliable performance while remaining computationally efficient.

These findings underscore the importance of aligning model selection with both performance objectives and operational constraints---especially in high-risk domains where minimizing undetected threats is critical. Our real-time prediction framework, for instance, completes the full decision cycle in approximately 9.1 seconds, including both static and dynamic feature extraction. The classification latency itself remains negligible (as low as 0.0001 seconds) when using a pre-trained XGBoost model. Even in scenarios involving runtime training, the performance overhead is marginal (9.10026 seconds total), and the model retains robust accuracy with a 4.56\% error margin.

\subsection{Runtime Performance Analysis}

On average, the extraction of static features—primarily opcode-based representations—required approximately 6.1 seconds per sample, including disassembly and opcode parsing via the Radare2 framework and its Python interface.

During the dynamic analysis phase, all samples were executed for varying durations to determine the minimum time required to collect sufficient features for machine learning model training. Based on these trials, three seconds per sample was identified as the minimum duration necessary to reliably monitor API calls and memory page protection flags. Each binary was executed using Frida within a sandboxed Windows environment, and dynamic traces were collected through lightweight instrumentation.

In addition, all extracted features were subjected to data preprocessing and dimensionality reduction, followed by machine learning model training and testing. The preprocessing and dimensionality reduction phase required approximately seven milliseconds per sample, broken down as follows: correlation calculation (6 ms), opcode counter processing (0.6 ms), and data scaling (0.1 ms).

Training and classification times for various machine learning models were presented in the previous section. Among them, the Soft Voting model required 127 milliseconds per sample for training and 9 milliseconds per sample for classification. 

As a result, the proposed framework is capable of detecting ransomware in approximately 9.2 seconds, comprising 9.1 seconds for feature extraction and analysis, and 0.1 seconds for data preprocessing and classification.

These results demonstrate that the proposed hybrid analysis pipeline is computationally efficient and scalable. Despite the common perception that dynamic analysis incurs higher computational overhead, the relatively low per-sample runtime observed in both analysis phases supports the framework’s applicability to large-scale batch processing and offline malware analysis workflows.

\section{Discussion}\label{sec:discussion}

The results presented in Section \ref{sec:results} confirm the effectiveness of our hybrid ransomware detection framework in many respects. However, beyond these experimental findings, we would like to discuss in this section the effectiveness, novelty, and open challenges of our proposed framework.

RansomTrack offers a modular and lightweight hybrid behavioral analysis pipeline that leverages static features at the opcode level and dynamic features, including detailed API call traces and memory page protection flags. RansomTrack provides instruction-level disassembly (via Radare2) and dynamic API hooking (via Frida) in six behavioral areas: memory operations, mutex interactions, registry manipulation, file activity, shell execution, and networking, along with memory page protection flags.

Unlike heavy-duty sandbox or kernel-level monitoring solutions, our framework relies solely on user-space tooling (Radare2 and Frida), providing features such as easy portability across various deployment environments, real-time applicability with sub-10-second decision latency, and interpretability through SHAP-based explanations.

The RansomTrack dataset, containing 165 ransomware families, has improved generalization across obfuscation, packaging, and polymorphic variations, making the proposed framework more robust than signature-based or narrowly scoped detection methods.

While existing systems such as RansomFormer and RanSMAP, which we review in detail in Section \ref{sec:related_work}, offer strong performance in isolad methods, they also have limitations such as lack of behavioral coverage such as lack of mutual exclusion/network monitoring, high computational cost or instrumentation complexity, and limited support for early stage detection.

RansomTrack addresses these gaps in the literature by integrating structural and runtime indicators, providing minimal operational overhead thanks to scriptable tools, and near-time decision-making capabilities with negligible classification latency.

Despite the promising results obtained, there are still challenges to be solved, such as the lack of a multi-window or event-triggered execution strategy for dynamic tracking, memory entropy features, and code similarity fingerprinting for load detection, and richer semantic modeling of behavioral traces.

Despite these challenges, RansomTrack's modularity and low latency provide the opportunity to be used in a variety of real-world scenarios. In particular, endpoint protection agents can use pre-trained models for local inference, security operations centers (SOCs) can integrate the system for forensic triage or alert verification, critical infrastructure environments can use ensemble-based classifiers like Soft Voting, and consumer-side endpoint protection solutions can opt for faster models like XGBoost. Furthermore, for analysts, SHAP-based feature interpretability supports explainable decision making.

\section{Conclusion}
\label{sec:conclusion}

In this paper, we introduce RansomTrack, a hybrid ransomware detection framework that integrates static opcode-level features with dynamic API-level and memory page protection behavioral signals. The static component extracts instruction-level semantics via Radare2, while the dynamic module uses Frida to monitor runtime behaviors in six key areas, including memory, registry, and file operations. By leveraging these tools, we have shown that high-accuracy, low-latency detection can be achieved without resorting to heavy sandboxing or kernel-level interventions. Unlike traditional approaches that rely on sandbox logs or byte n-grams, RansomTrack provides fine-grained, real-time instrumentation by capturing detailed API arguments and memory protection changes. This enables precise detection of behaviors such as unpacking, code injection, and evasion techniques.

Our evaluation on a dataset of 2,410 labeled PE files from 165 ransomware families showed that hybrid modeling significantly improves classification performance compared to static-only or dynamic-only approaches.

Among the tested classifiers, the Soft Voting ensemble provided the highest recall, while XGBoost offered the lowest inference latency, enabling deployment in both high assurance and real-time security contexts. The SHAP-based analysis further highlighted key distinguishing features.

Beyond the empirical gains, RansomTrack also addresses critical gaps in the literature, such as limited family diversity, lack of interpretability and runtime inefficiencies. It offers a fast, extensible and interpretable pipeline with an average runtime of 9.1 seconds per sample. Its modular design supports selective probe activation and customized feature recording, making it suitable for real-time applications.

However, challenges such as evasion-sensitive ransomware, progressive execution and encrypted payloads still remain open research issues.

Future work will investigate multi-stage dynamic execution, adaptive retraining mechanisms, and deep behavior modeling using graph-based or temporal learning techniques. We also plan to publish the RansomTrack dataset and tools as an open benchmark to facilitate reproducible research and cross-method comparisons in the field of ransomware detection.


\appendix
\section{Appendix: RansomTrack Dataset Information Card}

\subsection{Structure of the Dataset file}

\begin{table*}[b]
\caption{List of ransomware families and number of samples in the dataset.}
\label{tab:ransomware_appendix}
\centering
\scriptsize
\resizebox{\textwidth}{!}{%
\begin{tabular}{|l|c|l|c|l|c|l|c|l|c|l|c|}
\hline
\textbf{Ransomware Family} & \textbf{Count} & \textbf{Ransomware Family} & \textbf{Count} & \textbf{Ransomware Family} & \textbf{Count} & \textbf{Ransomware Family} & \textbf{Count} & \textbf{Ransomware Family} & \textbf{Count} & \textbf{Ransomware Family} & \textbf{Count} \\
\hline
Adhubilka & 2 & AESCRYPT & 1 & AESRT & 1 & AgentTesla & 2 & Alien & 1 & Avaddon & 51 \\
AvosLocker & 1 & AXLocker & 5 & Babadeda & 1 & Babuk & 47 & Bazek & 2 & BlackBasta & 23 \\
BlackCat & 53 & Blackmatter & 11 & BlackOut & 3 & BlackShades & 1 & BlackSnake & 1 & BlackSuit & 3 \\
BuerLoader & 3 & Bulkwar & 1 & Buran & 1 & Cerber & 37 & Chaos & 51 & Chimera & 2 \\
Cipher & 1 & CoinMiner & 1 & Conti & 54 & CryLock & 2 & CrypHydra & 1 & Crypt360 & 1 \\
CryptFile2 & 1 & CryptNet & 1 & CryptoJoker & 1 & CryptoLocker & 1 & CryptoWall & 1 & Crypute & 1 \\
CyberVolk & 5 & Cyborg & 1 & Cylan & 2 & Cylance & 2 & Dacic & 1 & Dalexis & 1 \\
Darkside & 5 & DarkWatchman & 1 & Erica & 1 & Eternity & 1 & EternityStealer & 1 & Exmas & 1 \\
Exorcist & 13 & Eyedocx & 1 & Fabiansomware & 1 & Formbook & 1 & Gandcrab & 36 & GarrantDecrypt & 1 \\
Genasom & 2 & GlobeImposter & 3 & GpCode & 2 & HakunaMatata & 2 & Hardbit & 2 & Heracles & 1 \\
Hermes & 1 & HiddenTear & 2 & Hitobito & 2 & Hive & 1 & HolyGhost & 1 & Huntr & 1 \\
HydraCrypt & 1 & Inlock & 1 & Kadavaro & 1 & KillMBR & 1 & Kryptik & 1 & levislocker & 1 \\
Levislocker & 1 & LimeRAT & 2 & LockBit & 134 & Locker & 1 & LockerGoga & 1 & Locky & 7 \\
LokiLocker & 2 & Lynx & 2 & MafiaWare666 & 1 & Makop & 20 & Mallox & 6 & Mammon & 5 \\
MassLogger & 1 & Matrix & 1 & Maze & 8 & MedusaLocker & 2 & Mimic & 1 & Mimikatz & 2 \\
Mole & 1 & MoneyMessage & 3 & Mountlocker & 3 & Mydoom & 2 & Nabucur & 3 & Nefilim & 18 \\
Neshta & 2 & Netwalker & 32 & Nitro & 3 & Njrat & 1 & PenterWare & 2 & Phalcon & 1 \\
Phobos & 43 & PLAY & 5 & PlutoCrypt & 1 & Pony & 1 & Prestige & 1 & PureCrypter & 1 \\
PureLogStealer & 1 & Pysa & 38 & Qakbot & 1 & QuasarRAT & 3 & RagnaLocker & 1 & Ragnarok & 35 \\
RansomeXX & 3 & Redeemer & 2 & RedLineStealer & 1 & ReturnBack & 1 & Royal & 1 & RustyStealer & 1 \\
Ryuk & 37 & Ryzerlo & 2 & Sage & 2 & Saturn & 1 & Scar & 1 & Seven & 3 \\
ShadowRoot & 1 & Slam & 1 & Smert & 1 & Spacecolon & 1 & StayHigh & 1 & Stop & 51 \\
StormKitty & 1 & Surtr & 3 & TankRansom & 2 & TankixCrypt & 2 & Targeted & 4 & TeslaCrypt & 43 \\
Thanos & 34 & TrickBot & 2 & Trigona & 4 & Troldesh & 1 & Venus & 3 & Virlock & 1 \\
Vohuk & 1 & VoidCrypt & 1 & WannaCry & 5 & Wastedlocker & 10 & Xorist & 1 & Xworm & 2 \\
Yanluowang & 1 & Zeppelin & 10 & Zerber & 6 & Zusy & 1 & & & \\
\hline
\end{tabular}
}
\end{table*}

The dataset is presented in \texttt{csv} file format and contains behavioral and structural features extracted from a total of 2,410 executable samples, including both ransomware and benignware. 

The first 1,205 rows represent ransomware samples, while the remaining 1,205 rows correspond to benign software. The dataset is organized as follows:

\begin{itemize}
    \item \textbf{Column 0:} Contains the \texttt{SHA-256} hash values of the files, serving as unique identifiers.
    \item \textbf{Columns 1--1002:} Represent static \texttt{opcode} frequency features extracted from the binary code of each sample.
    \item \textbf{Columns 1003--2737 and 2742--6027:} Contain dynamic \texttt{API call} frequency features collected during runtime execution using dynamic analysis tools.
    \item \textbf{Columns 2738--2741:} Encode memory protection behaviors associated with runtime execution, such as \texttt{PAGE\_READWRITE}, \texttt{PAGE\_EXECUTE\_WRITE}, and other memory-related flags.
    \item \textbf{Column 6028:} Labeled as \texttt{label}, this column indicates the ground-truth class of the file, where \texttt{0} denotes benignware and \texttt{1} denotes ransomware.
\end{itemize}

This feature layout provides a comprehensive representation of both static and dynamic behaviors, enabling hybrid analysis and classification tasks.

\subsection{Ransomware Families and Sample Counts}

The ransomware families and their sample numbers within the dataset are presented in Table~\ref{tab:ransomware_appendix}.

\clearpage

\printcredits

\bibliographystyle{cas-model2-names}

\bibliography{cas-refs}

\bio{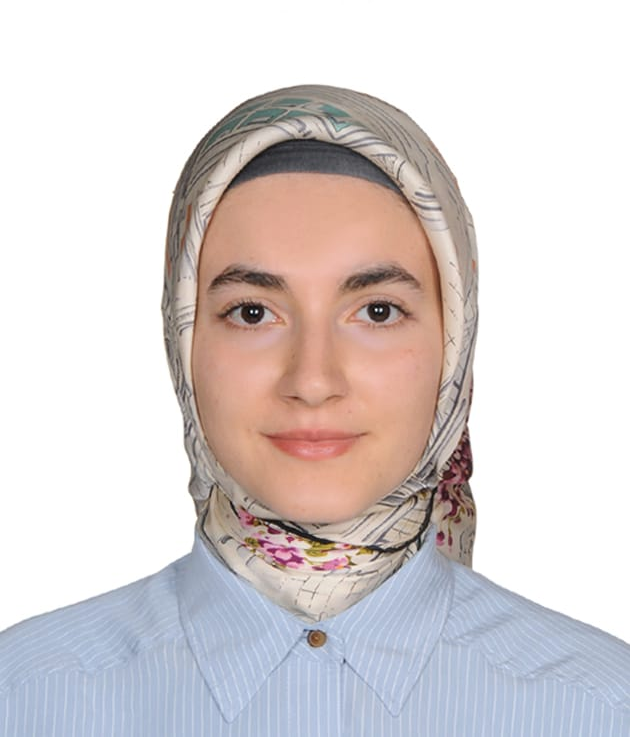}
Busra Caliskan received the B.S. degree in biomedical engineering from Yeditepe University, Istanbul, Turkiye, in 2020, and the M.S. degree in computer engineering from Istanbul Technical University (ITU), Istanbul, in 2025. She is currently pursuing the Ph.D. degree in computer engineering at ITU.
\endbio

\bio{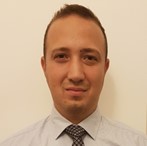}
Ibrahim Gulatas received the B.S. degree in computer engineering from the Turkish Naval Academy, Istanbul, Turkey, in 2010. He received his M.S. degree in computer engineering from Bahcesehir University, Istanbul, Turkey, in 2018, and his Ph.D. in Computer Engineering from Istanbul Commerce University, Istanbul Turkey, in 2023.  His current research interests include information security, malware analysis, and machine learning applications. 

He has been a navy officer in the Turkish Naval Forces since 2010. His current position in the Navy is at the National Defense University in Istanbul.
\endbio

\bio{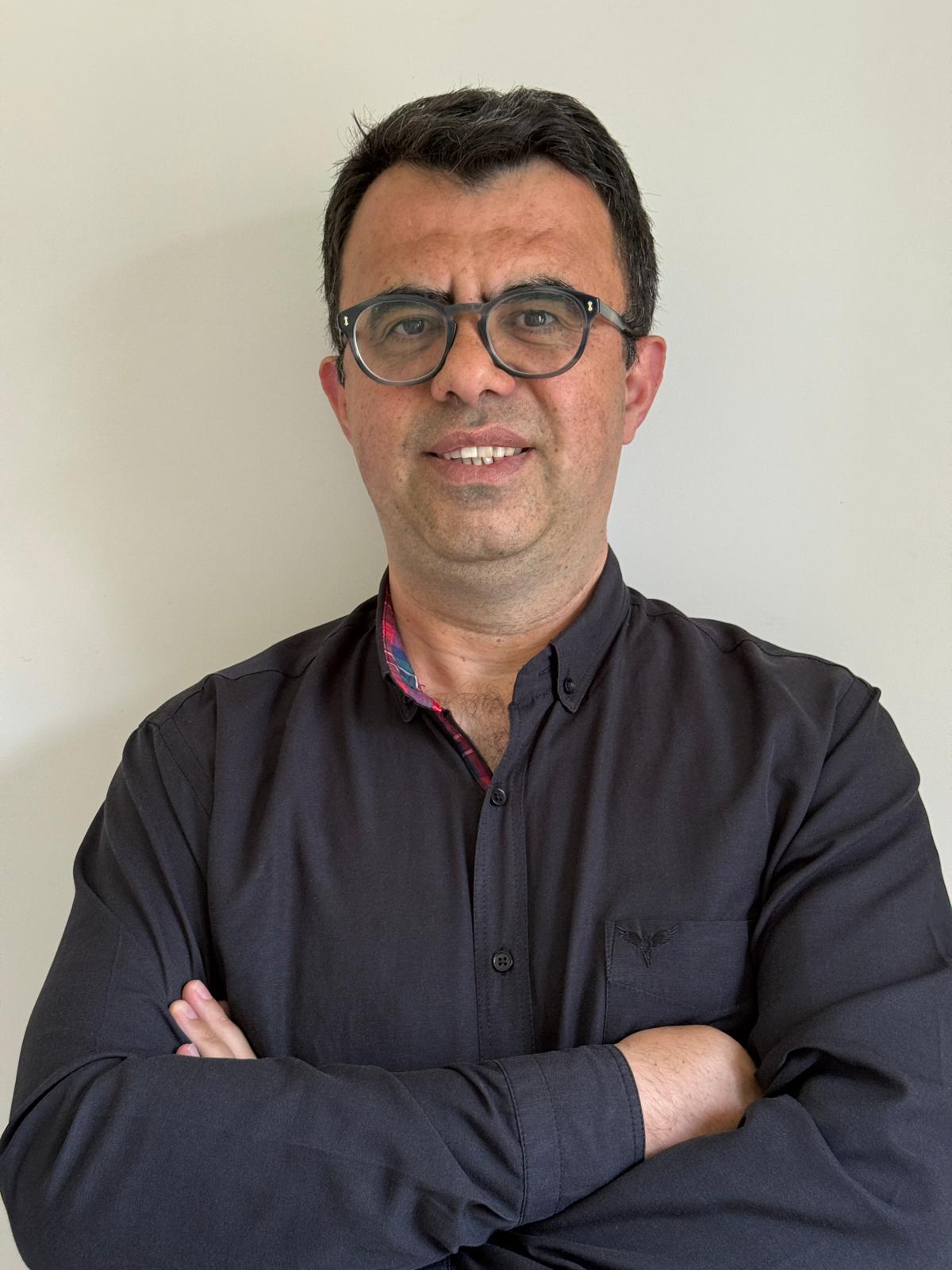}
H. Hakan Kilinc received his B.S. degree in Mathematics and Computer Science from Ege University, Izmir, Turkey, in 1997. He received his M.S. degree in 2001 in the Department of Computer Engineering at the Izmir Institute of Technology, Izmir, Turkey. He holds a Ph.D. about the security of SIP (Session Initiation Protocol) from the Department of Electronics Engineering at the Gebze Technical University in 2014.

He was a visiting scholar at the University of Texas at Dallas from 2009 to 2011, a cybersecurity product line manager for Netas from 2014 to 2019, and, since December 2019, an Innovation Project Manager for Orion Innovation Turkey in Istanbul, Turkey.   
\endbio

\bio{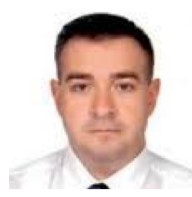}
A. Halim Zaim received his B.S. degree in computer engineering from Yildiz Technical University, Istanbul, Turkey, in 1993. He received his MS degree in Computer Engineering from Bogazici University, Istanbul, Turkey in 1996 and his Ph.D. in Electrical and Computer Engineering from North Carolina State University (NCSU), NC, USA, in 2001. His research interests include the IoT, big data, network design, cyber security, network security, and communication network protocols.

He was the Vice-Rector of Istanbul Commerce University. He is currently Dean with the Department of Computer Engineering, Faculty of Engineering, Istanbul Technical University, Istanbul, Turkey.
\endbio

\end{document}